\documentclass[useAMS]{mn2e}

\def\kms{$\rm{km\,s^{-1}}$}
\def\spose#1{\hbox to 0pt{#1\hss}}
\def\lta{\mathrel{\spose{\lower 3pt\hbox{$\mathchar"218$}}\raise 2.0pt\hbox{$\mathchar"13C$}}}
\def\gta{\mathrel{\spose{\lower 3pt\hbox{$\mathchar"218$}}\raise 2.0pt\hbox{$\mathchar"13E$}}}

\title[PKS1932-464: a radio source in an interacting
  group?]{PKS1932-46: a radio source in an interacting
  group?\thanks{Based on observations collected at the European
    Southern Observatory, Chile (program 075.B-0820(A) and 078.B-0500(A)).}}
\author[K.\,J. Inskip et al ]{K. J. Inskip$^{1}$\thanks{E-mail:
k.inskip@shef.ac.uk}, C. N. Tadhunter$^{1}$, D. Dicken$^{1}$,
 J. Holt$^{1}$, M. Villar-Mart\'{i}n$^{2}$, \newauthor R. Morganti$^{3}$, 
 \\
$^{1}$ Department of Physics \& Astronomy, University of Sheffield, Sheffield S3 7RH\\
$^{2}$ Instituto de Astrof\'{i}sica de Andaluc\'{i}a (CSIC),
  Aptdo. 3004, 18080 Granada, Spain\\
$^{3}$ Netherlands Foundation for Research in Astronomy, Postbus 2,
  7990 AA Dwingeloo, The Netherlands
}
\begin{document}

\date{}

\pagerange{\pageref{firstpage}--\pageref{lastpage}} \pubyear{2007}

\maketitle
\label{firstpage}

\begin{abstract}We present the results of a multiwavelength study of
the $z = 0.23$ radio source PKS1932-46.  Integral field unit
spectroscopy using the Visible Multiobject Spectrograph (VIMOS) on the
VLT is
used to study the morphology, kinematics and ionisation state of the
extended emission line region (EELR) surrounding this source, and also
a companion galaxy at a similar redshift.  Near- and far-infrared
imaging observations obtained using the NTT and SPITZER are used to
analyse the underlying galaxy morphologies and the nature of the AGN. 

The host galaxy is identified as an $\sim$M$_\star$
elliptical. Combining Spitzer mid-IR with X-ray, optical and near-IR
imaging observations of this source, we conclude that its AGN is
underluminous for a radio source of this type, despite its status as a
broad-line object.  However, given its relatively large
[O\textsc{iii}] luminosity it is likely that the AGN was substantially
more luminous in the recent past ($\lta 10^4$ years ago).

The EELR is remarkably
extensive and complex, reminiscent of the systems observed around
sources at higher redshifts/radio powers, and the gas is predominantly ionised by a
mixture of AGN photoionisation and emission from young stars. We
confirm the presence of a series of star-forming knots extending
north-south from the host galaxy, with more prodigious star formation
occuring in the merging companion galaxy to the northeast, which has
sufficient luminosity at mid-- to far--IR wavelengths to be classified
as a luminous infrared galaxy (LIRG).

The most plausible explanation of our observations is that PKS1932-46
is a member of an interacting galaxy group, and that the impressive EELR is
populated by star-forming, tidal debris. We suggest that the AGN
itself may currently be fuelled by material associated either with the
current interaction, or with a previous merger event.  
Surprisingly, it is
  the companion object, rather than the radio source host galaxy,
  which is undergoing the bulk of the star formation activity within
  the group. 
\end{abstract}


\section{Introduction}

Extended emission line regions (EELRs) are often observed around
powerful distant radio sources (McCarthy et al 1987), and their observed
properties (size, luminosity, kinematics and ionisation state) are
known to depend strongly on those of the radio source (e.g. Best et al
2000, Inskip et al 2002a, Moy \& Rocca-Volmerange 2002).  
Although the behaviour of the  emitting material and the balance
between different ionisation mechanisms is now becoming increasingly
well understood, many outstanding questions still remain.  The origin
of the extended emission line gas is one such issue.  Does it exist
{\it in situ} prior to the 
onset of the radio source? Is it formed from material driven out of  the
host galaxy by AGN/radio source/starburst related outflows? Is it
infalling material associated with a cooling flow? Or, is it
produced via galaxy interactions/mergers associated with the radio
source triggering event?  A secondary issue is whether the distribution
of material is related to its origin, and its eventual fate. As the
properties of this gas may have close links with both the origin of
radio source activity and the 
host galaxy evolution, these are particularly pertinent questions.

While the emitting material is most frequently observed either along
the radio source axis or within AGN ionisation cones, recent
observations have shown that this is not always the case. Extensive
emission line regions lying almost perpendicular to the radio axis
have been identified around several radio sources, via recent narrow band imaging observations
(Tadhunter et al 2000, Sol\'{o}rzano-I\~{n}arrea et al 2002,
Villar-Mart\'in et al 2005).  These features may perhaps be related to the
large gaseous haloes observed at  higher redshifts (e.g. R\"{o}ttgering
et al 1995, van Ojik et al 1997, Jarvis et al 2003, Villar-Mart\'in et
al 2003), but their exact nature is still an open question.
Large, complex EELRs with no clear-cut link with the radio source morphology
have also been observed around low redshift quasars, e.g. 4C 37.43 (Fu
and Stockton 2007, Stockton et al 2002).

The EELR surrounding the FRII radio galaxy PKS1932-464 ($z \sim 0.231$)
is a prime example of such highly complex systems.   Previous imaging and spectroscopic
observations of this source (Villar-Mart\'{i}n et al 1998,
Villar-Mart\'{i}n et al 2005; hereafter VM98 and VM05) have shown that 
the host galaxy is surrounded by an extensive ($\sim 100$kpc radius), knotty emission line
region, extending well beyond the observed radio source. As well as
material lying along the radio source axis, emission is also 
observed from structures lying at large angles to the radio source axis,
which are unlikely to lie within the AGN ionisation cone. 
VLT long-slit spectroscopy at position angle PA -9, misaligned by
$63^\circ$ from the 
radio source axis (PA -72),
 has determined that this off-axis emission is likely to be due to
star-forming objects (VM05). 

Extreme EELR features such as those displayed in the case of
PKS1932-464 (extreme size, chaotic morphology of the ionised gas,
off-axis star-forming knots and radio source asymmetries) are
generally more common at higher redshifts than they are at lower
redshifts (particularly in terms of
EELR morphologies and kinematics; Inskip et al 2002b). Similar trends with
redshift are observed for the continuum emission from these regions
(the {\it Alignment effect}; Chambers, Miley \& van Breugel 1987;
McCarthy et al. 1987; Allen et al 2002).    In part, these trends are
due to an increased incidence of jet-cloud interactions at higher
redshifts, but environmental factors may also play an important role.
At lower redshifts the general trend is for smaller, more regular EELRs
which are usually (but not always) well-aligned with the radio source axis. 
Previous spectroscopy of PKS1932-464 has not isolated the extreme
kinematics typical of jet-cloud interactions, and the complex
distribution of star-forming EELR material (a rarity at such low redshifts) and the proximity of a
possible companion galaxy are suggestive that galaxy interactions and
environmental factors are the primary cause of the observed EELR
features.  

One option for exploring these systems in detail is the use of
integral field spectroscopy (IFS).   IFS studies of EELRs have the
potential to greatly improve our understanding of these systems,
particularly in terms of building up a consistent explanation for the
nature, distribution and origin of the emission line gas, and the
links with the radio source triggering mechanism. The additional
spatial data provided by IFS observations is a major advantage, as it
allows the EELR properties (physical conditions, gas dynamics,
ionisation state) to be efficiently studied and quantified as a
function of  position relative to the radio source, rather than just
along the radio axis. Most previous long slit studies have concentrated
on the radio axis position angle, and are therefore biased towards the
regions for which the gas kinematics are perturbed by the growing
radio source rather than reflecting the intrinsic properties of the
extended halo.  

In this paper, we present the results of a
multi-wavelength study of PKS1932-464, combining optical IFS
spectroscopy with near and far infrared imaging observations, and
allowing us to greatly expand on the previous work carried out on this
source (VM98, VM05). The
details of the source, our
observations and data reduction are described in section 2, and the results
presented in section 3, including an analysis of the properties of the
galaxy previously  assumed to be a nearby companion.  In section 4, we
discuss the implications of our results, and we present our
conclusions in section 5.   Throughout this paper, we assume
cosmological parameters of $\Omega_0 = 0.3$, $\Omega_{\Lambda} =
0.7$ and $H_0 = 70 \rm{km\,s^{-1}\,Mpc^{-1}}$, which result in  an
angular scale of $3.7$kpc/arcsec at $z = 0.23$. 

\section{Source details, observations and data reduction}

\subsection{PKS1932-46}

As described in the introduction, the EELR surrounding PKS1932-46 is
quite remarkable for a low redshift radio galaxy.  However, this
paper also considers the wider properties of the host galaxy and radio
source in addition to the EELR, and it is therefore appropriate to
briefly outline the other key properties of this source.

PKS1932-46 is an FRII radio galaxy lying at a redshift of $z \sim
0.231$ with a total radio power of $5.5\times 10^{26} \rm W Hz^{-1}$ at
$\sim 5$GHz (Morganti, Killeen \& Tadhunter 1993). Its radio structure
(VM98; see also Fig~\ref{Fig: 1}) is 
somewhat asymmetric; the western radio lobe has a projected physical
extent of 41kpc cf. 26kpc for the eastern radio lobe.  Core emission
is observed weakly at 2.3 and 8.6GHz (Morganti et al 1997; VM98), and
the source has an R-parameter of $R_{\rm 2.3 GHz} = 0.0023$. In terms
of polarisation, the western lobe appears more strongly polarised than
the eastern lobe,
with strong depolarisation of the eastern lobe at lower radio
frequencies (VM98), possibly due to warm emission line gas in this region.

\subsection{VIMOS IFU spectroscopy}

Integral field unit (IFU) spectroscopic observations were carried out using the Visible
Multiobject Spectrograph (VIMOS; Le F\'{e}vre et al 2003, Scodeggio et
al 2005), on 2005 June 4, 2005 June 7, and 2005 June 8, as part of the
European Southern Observatory (ESO) observing programme
075.B-0820(A). 
VIMOS is mounted on the Nasmyth focus of the UT3 Melipal unit of the
Very Large Telescope (VLT).  Our data were obtained using the medium
resolution (MR-orange) grism and the GG475 order sorting filter; in this configuration, the IFU consists
of 1600 microlenses coupled to 0.67-arcsec diameter fibres, covering a
total sky area of $27 \times 27$ arcsec$^2$, with a useful wavelength
range of $\sim 4500-9000$\AA. The instrumental spectral resolution, as
determined from unblended skylines, was approximately 7\AA\ (FWHM).  The
total integration time was $\sim 12450$s; full details of the
observations and observing conditions are given in Table 1.  The data
were obtained with two separate pointings, giving an overall field of
view with a maximum extent of 41 arcseconds.  As well as allowing a
larger region of the EELR and additional objects to be studied, large
offsets between exposures should ideally have facilitated accurate sky subtraction.

\begin{figure}
\vspace{5.9 in}
\begin{center}
\includegraphics{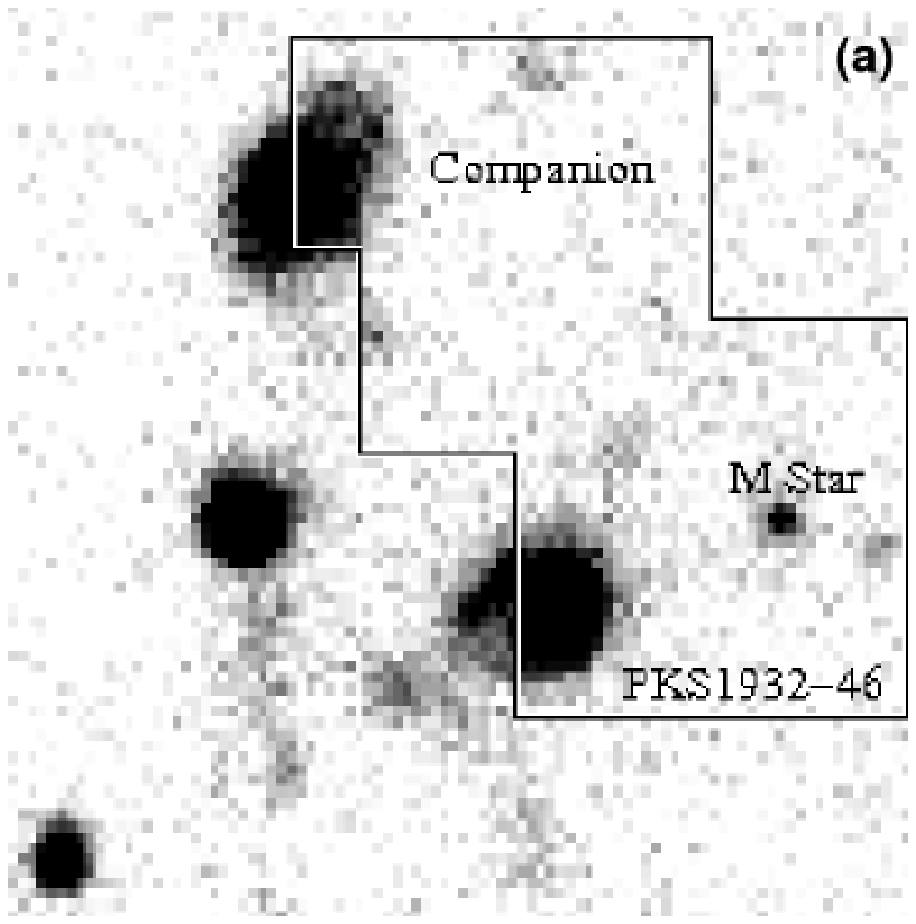}
\includegraphics{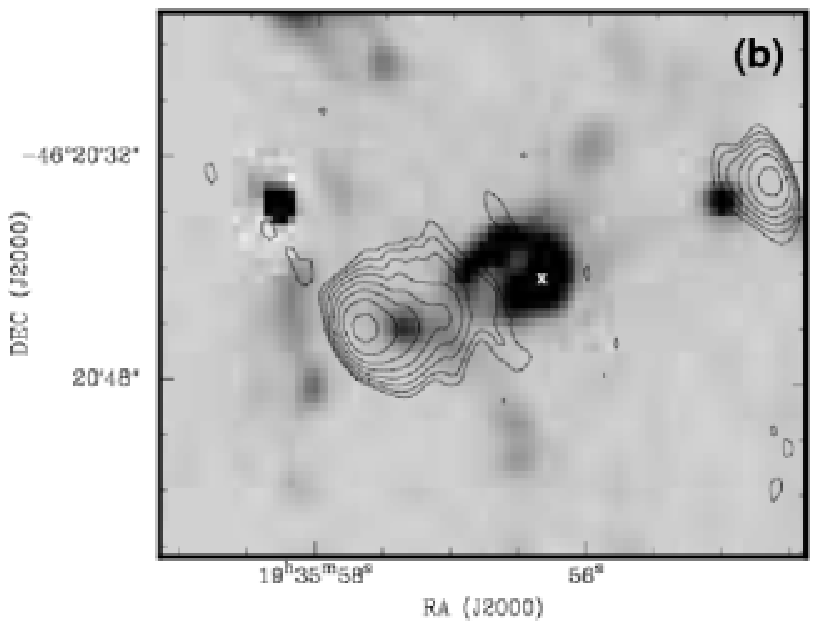}
\includegraphics{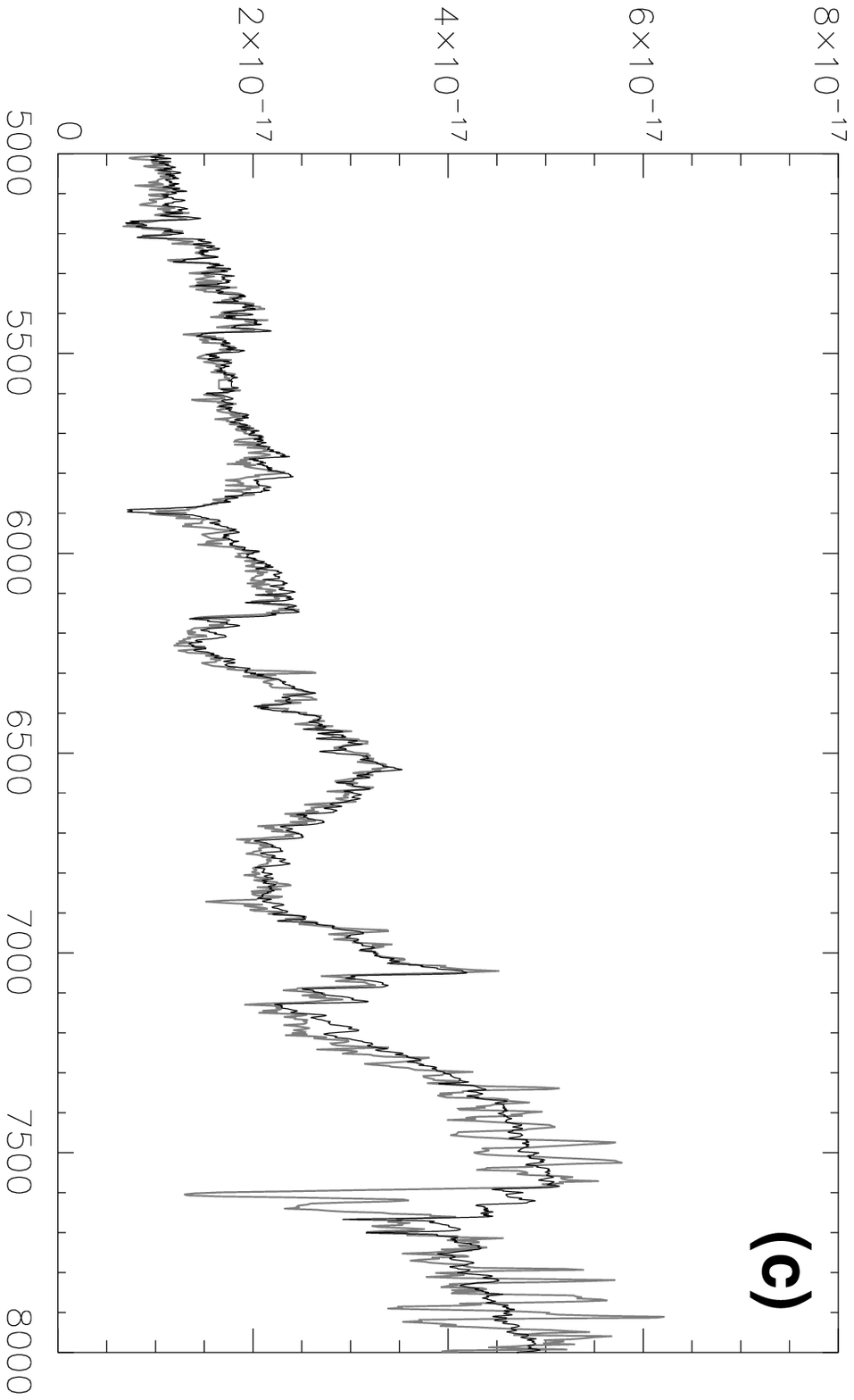}
\end{center}
\caption{(a - top) [O\textsc{ii}]+continuum image of PKS1932-46 and the
  surrounding field (from VM98),
  illustrating the approximate IFU field of view and identifying individual
  objects.  (b - centre) 5.8 GHz radio contours overlaid on an
  [O\textsc{ii}]+continuum image (from VM98),
  illustrating the relative size of the radio source.  (c - bottom)
  Extracted spectrum of the M star lying to the west of 
  PKS1932-464 (dashed line), 
  overlaid with a linearly scaled 50\% mixture  of 
  M2 and M3 SDSS stellar cross-correlation 
  template spectra (solid line).  
\label{Fig: 1} }
\end{figure}

Data reduction was carried out using the VIMOS Interactive Pipeline
and Graphical Interface (\textsc{vipgi}) software package (Scodeggio
et al 2005, Zanichelli et al 2005); here we summarise the basic steps
involved, which include some variations from the standard
\textsc{vipgi} procedures.  In the first stage of the reduction
process, bias subtraction and flat fielding is carried out separately
for the four IFU quadrants. For each target/spectrophotometric
standard star observation, the locations of the
spectra for each individual fibre are identified and traced using continuum lamp exposures.  Once all spectra
locations have 
been accurately traced, the spectra are wavelength calibrated using
arc-lamp observations obtained using the same instrumental set-up,
with resulting residuals typically $\sim 0.3$\AA\ per fibre on average.  The 
third stage of the reduction process is the extraction of individual
spectra.  Each individual fibre spectrum is extracted in two dimensions,
converted to a linear wavelength scale, and summed spatially to form a
one-dimensional spectrum. Cross-talk effects due to 
fibre--to--fibre contamination are likely to be minimal (uncertainties
of the order of $\sim5\%$); as any corrections
would introduce uncertainties of a similar magnitude (Zanichelli et al
2005), no corrections are made for this effect.  Cosmic rays were detected and removed from
the data using a 6$\sigma$ clipping algorithm evaluated within a 40
pixel box.  The instrumental sensitivity function for each IFU
quadrant was determined using the summed extracted spectra of
spectrophotometric standard star frames; this is then applied to the
science frames to produce spectra with accurate relative flux
calibration.  The data from 2005 June 7 and 2005 June 8 were obtained
in non-photometric conditions, and were flux calibrated via reference to the
photometric data obtained on 2005 June 4.

\begin{table}
\caption{Details of the VIMOS and SOFI observations and observing
  conditions. The seeing measurements are given for the wavelength
  of the observations.
}
\begin{tabular}{ccccc}
Date & Instrument &Exposure  & Seeing & Photometric?\\
& & time (s) & \\\hline
20050604 & VIMOS & 4980 & 0.3-0.7$^{\prime\prime}$& Fully\\
20050607 & VIMOS & 4980 & 0.8-1.2$^{\prime\prime}$& Partially\\
20050608 & VIMOS & 2490 & 0.6-0.7$^{\prime\prime}$& Partially\\
20061114 & SOFI & 3000 & 0.8$^{\prime\prime}$& Fully \\
\end{tabular}
\end{table}

Analysis of the night sky emission lines was used to correct for variations
in fibre-to-fibre transmission.  At this stage we also used the locations of six
bright unblended skylines to manually apply a corrective linear shift
in wavelength for each observation, for the minority of fibres
affected by this common fault.  An average sky spectrum can also
be produced and subtracted from the data.  However, for observations
of targets with extensive emission line regions, this step has an
unfortunate tendency to oversubtract flux in the vicinity 
of bright emission lines, even when the majority of the field of view
covers blank sky.  Pipeline sky subtraction was therefore not applied to our
data.  At this stage, we instead created three different versions of each
observed frame.  In the first version, we treat each frame as if
it were long-slit data: the fibres which include emission from our
target and other objects are identified (using the final data cube; by necessity,
this process is iterative in nature) and masked, and the remaining
fibres were used to derive a sky spectrum which is subtracted from the
data. This is effectively the same method as that used by the pipeline
for sky subtraction, but with the advantage that the relative flux
levels of emission line regions are not corrupted.  For our second
data set, we subtracted from each frame the data taken with the
alternative pointing position obtained immediately before/after the
frame in question.  This removes the sky flux and also gives an
accurate fringing correction, but where emission from different
objects lies at the same fibre position in both pointings, object flux
subtraction also occurs.  However, comparison with our other data sets
allows the affected fibres to be very clearly identified. 
Finally, we also retained an unaltered data set which included the
full sky emission, in order to have a means of identifying any
spurious features introduced by our treatment of the previous data
sets.  

The final stage of the reduction process was the creation of a
datacube, combining the four quadrant frames together and median
averaging each observation. As our data were obtained over several
nights (with different calibration data and observing conditions for
each), we initially kept the data for different nights and the two
pointing positions separate.  Fringe correction is an optional part of
this process, but was not included in our reduction as it cannot
accurately distinguish between the excess flux of fringing and
spatially extended line emission.  Although the VIMOS pointing is
accurate within a given set of observing blocks, this accuracy is not
maintained between observations made on different nights. The creation
of our final data cubes included one last step: registration and shifting
of the individual cubes based on the
position of bright point sources in the IFU field of view.

Fig.~\ref{Fig: 1} displays two [O\textsc{ii}]3727\AA+continuum images
of the field of PKS1932-46 (from VM98), the
first illustrating the field of view of our IFU data, and the second
showing the relative extent of the radio emission.    Our data, obtained with two separate pointings, include the host
galaxy, the EELR to the west, and the companion object to the
northeast. The somewhat unusual shape of our final FOV is caused by a block of faulty
fibres within the southeast quadrant of the IFU.   Extracted IFU
spectra confirm that the galaxy to the north-east lies at a redshift
of $z=0.2298 \pm 0.0002$, similar to that of the radio source itself
($z=0.2307 \pm 0.0002$). The bright 
object lying within the western radio lobe is an M2/3 star; in
Fig.~\ref{Fig: 1} we also display a comparison between M2/M3 SDSS stellar
cross-correlation template spectra (Stoughton et al 2002) and our
extracted spectrum for this object, confirming its nature as a
foreground star.

\subsection{SOFI Ks-band imaging}

\begin{table}
\caption{Details of our Spitzer MIPS and SOFI $K_S$-band photometry of
PKS1932-46 and the surrounding field. Columns 1 and 2 list the
wavebands and total exposure times.  Columns 3 and 4 give the 
extracted fluxes of PKS1932-46 and the nearby spiral companion
galaxy.  The $K_S-$band fluxes were evaluated within circular
apertures with 5 and 12 arcsecond radii, and the MIPS data are total
object fluxes derived from measurements of the PSF over various
apertures (typically 3-20 arcsecond radii). The 160$\mu$m flux of
PKS1932-46 is a 3-sigma upper limit.}
\begin{tabular}{ccr@{$\pm$}lr@{$\pm$}l}
Waveband  & Exposure & \multicolumn{4}{c}{Observed Flux (mJy)}\\
& time (s) &\multicolumn{2}{c}{PKS1932-46} &
\multicolumn{2}{c}{Companion} \\\hline
2.16$\mu$m ($5^{\prime\prime}$) & 3000.0s  & 0.60 & 0.02 & 1.38 & 0.03\\
2.16$\mu$m ($12^{\prime\prime}$) & 3000.0s  & 0.64 & 0.02 & 1.61
& 0.03\\
24$\mu$m &  180.4s &  2.5 & 0.1 & 4.4 & 0.4 \\
70$\mu$m &  545.3s & 17.6 & 2.4 & 48.3 & 5.1\\
160$\mu$m &  167.8s & \multicolumn{2}{c}{$<94.4$} & 198.1 & 31.5\\
\end{tabular}
\end{table}

Ks-band imaging observations of PKS1932-46 were obtained on 2006
November 14
(see Table 1 for details).
Fifty 1-min exposures 
were obtained using the SOFI instrument 
(Moorwood, Cuby \& Lidman 1998) on
the ESO 3.5-m New Technology Telescope (NTT).  The instrument was used
in the small field mode, giving a plate scale of 0.144 arcsec per
pixel, and each observation was subject to a random offset within a 40
arcsec diameter box, resulting in a typical field of view of $\sim 3 \times
3$ arcmin.  The data were corrected for SOFI's interquadrant row cross 
talk effect using an adapted version of the SOFI crosstalk.cl IRAF
script.  The data were flat-fielded using the following process: all target
frames were combined, median filtered and normalised to a mean pixel
value of 1.0 to create a first-pass flat-field
image, which was then applied to each frame.  Bright objects on the flat-fielded images
were then masked out, and the process repeated with the masked frames, allowing the data
to be cleanly flat fielded without any contamination from stars or
galaxies. 
The flat-fielded data were sky-subtracted and combined
using the IRAF package DIMSUM, creating a final mosaiced image of
approximately $180 \times 180$ arcsec$^2$, which was flux calibrated
using observations of NICMOS Photometric Standard stars (Persson et al
1998), giving a photometric zero-point magnitude of $K_S = 22.347 \pm
0.010$.  The data were corrected for galactic extinction using 
$E(B-V)$ values for the Milky Way from the NASA Extragalactic
Database (Schlegel et al 1998), and the parametrized
galactic extinction law of Howarth (1983).


\subsection{Spitzer MIPS photometry}

Spitzer MIPS (Werner et al 2004; Rieke et al 2004) photometry of
PKS1932-46 was carried out as part of a program of observations
of a large sample of 2Jy radio galaxies (see Tadhunter et al 1993 for
sample definition).   The Spitzer observations were made on
2006 May 5 in each of the 24$\mu$m, 70$\mu$m and 160$\mu$m wavebands;
details of the 
observations are provided in Table 2.
The data were reduced using the MIPS pipeline (Gordon et al 2005, Masci
et al 2005) and the MOPEX  mosaicing software (Makovoz \&
Khan 2005, Makovoz \& Marleau 2005) provided by the 
Spitzer Science Center. In addition to the standard mosaicing procedure, we
have also used a column filtering process on the 70um data (obtained
from the Spitzer Science Center's contributed software resource), and an overlap
correction program on the the 24$\mu$m data (available on the Spitzer
Science Center website).

\section{Results}

\begin{figure}
\vspace{2.4 in}
\begin{center}
\includegraphics{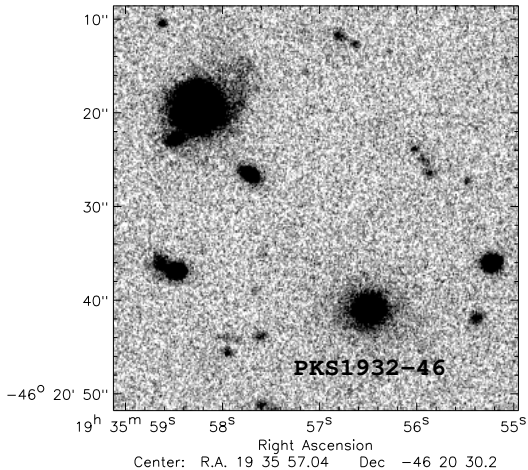}
\end{center}
\caption{SOFI $Ks-$band image of PKS1932-464 (lower right, labelled) and the
  surrounding field.
\label{Fig: kband}}
\end{figure}
\subsection{The host galaxy}

Figure \ref{Fig: kband} displays our new $Ks-$band imaging
observations of PKS1932-46 and its surrounding field. The radio source
host galaxy lies towards the bottom of the field, and is a fairly
regular elliptical. It has a total $K_S-$band magnitude (Galactic
extinction corrected and evaluated within a 10 arcsecond/37 kpc diameter
aperture) of $15.11 \pm
0.03$.
 The two sources at declinations $\sim -46^{\circ}
20^{\prime} 36^{\prime\prime}$ are both point sources; the western
object (R.A. $\sim 19^h 35^m 55^s.2$) is known to be an M2/M3 star,
while the eastern object (R.A. $\sim19^h 35^m 58^s.6$) lies adjacent
to some more diffuse emission to the east.  The large object to the
north-east of the field with a similar redshift to PKS1932-46 is a disturbed galaxy with clear signs of a
double nucleus and a tidal tail; this object has a 10
arcsecond diameter aperture magnitude of $14.21 \pm 0.02$ in the
$K_S-$band, and will be discussed in greater depth in section 3.3.

\begin{figure*}
\vspace{2.05 in}
\begin{center}
\includegraphics{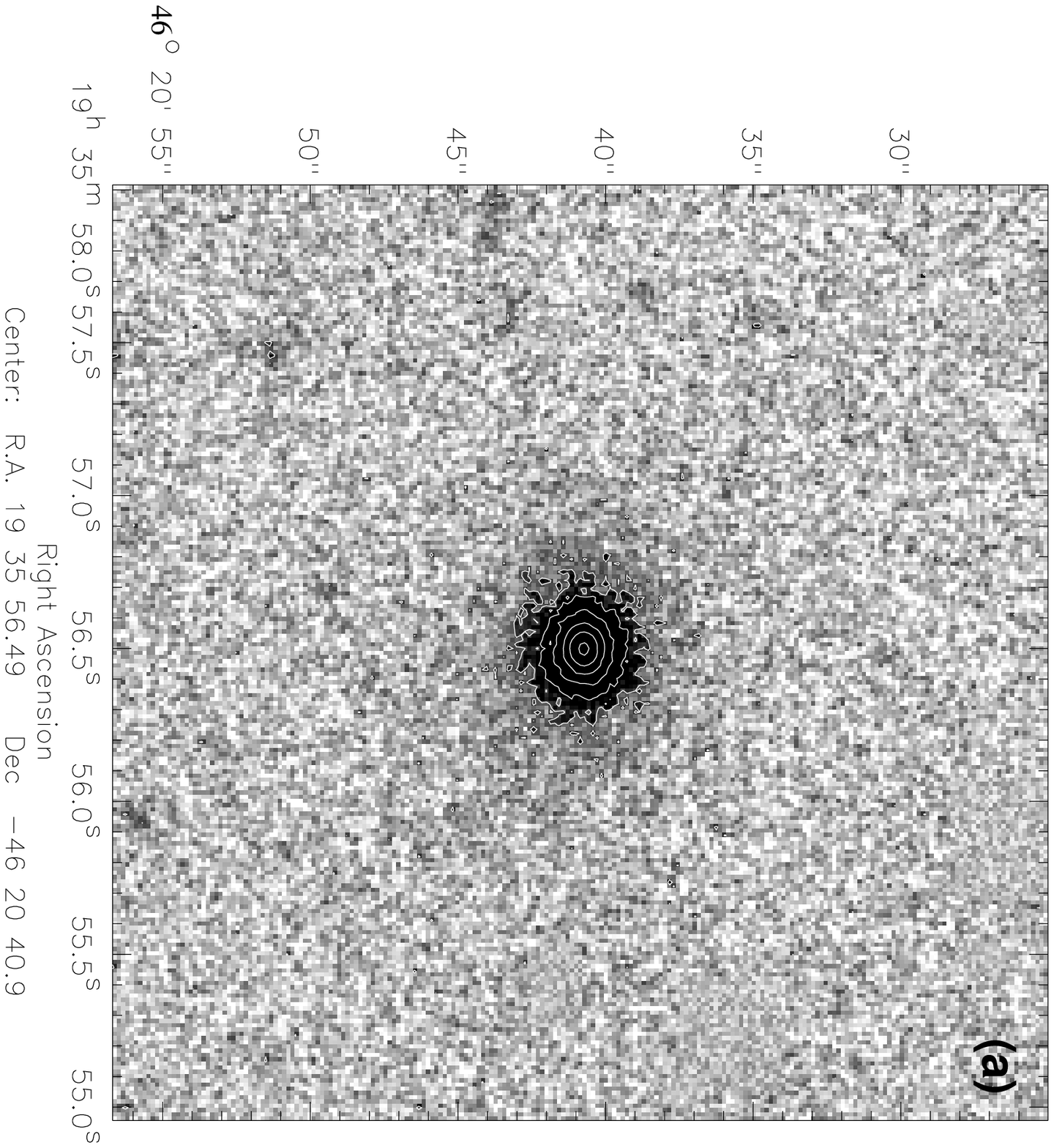}
\includegraphics{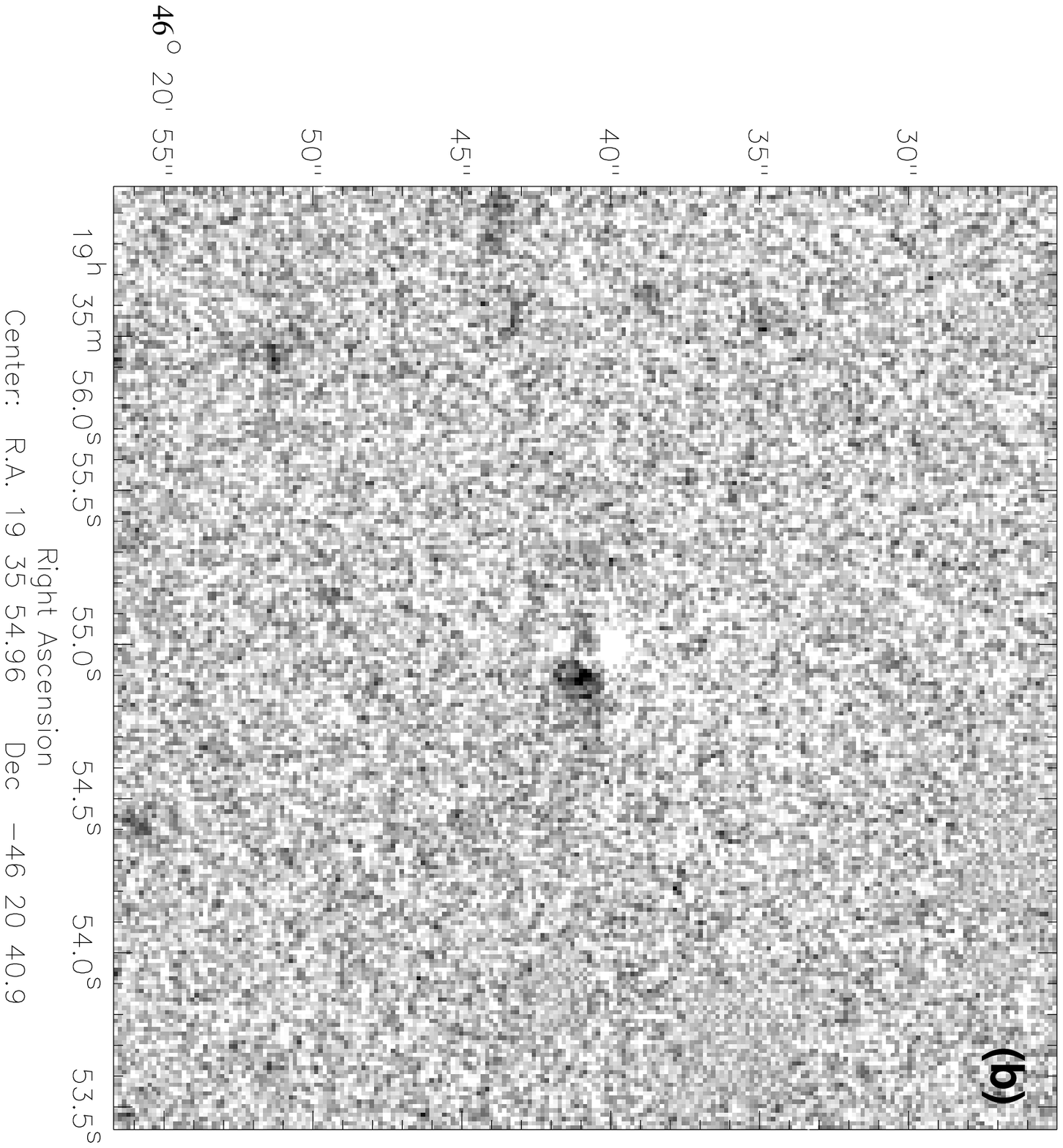}
\includegraphics{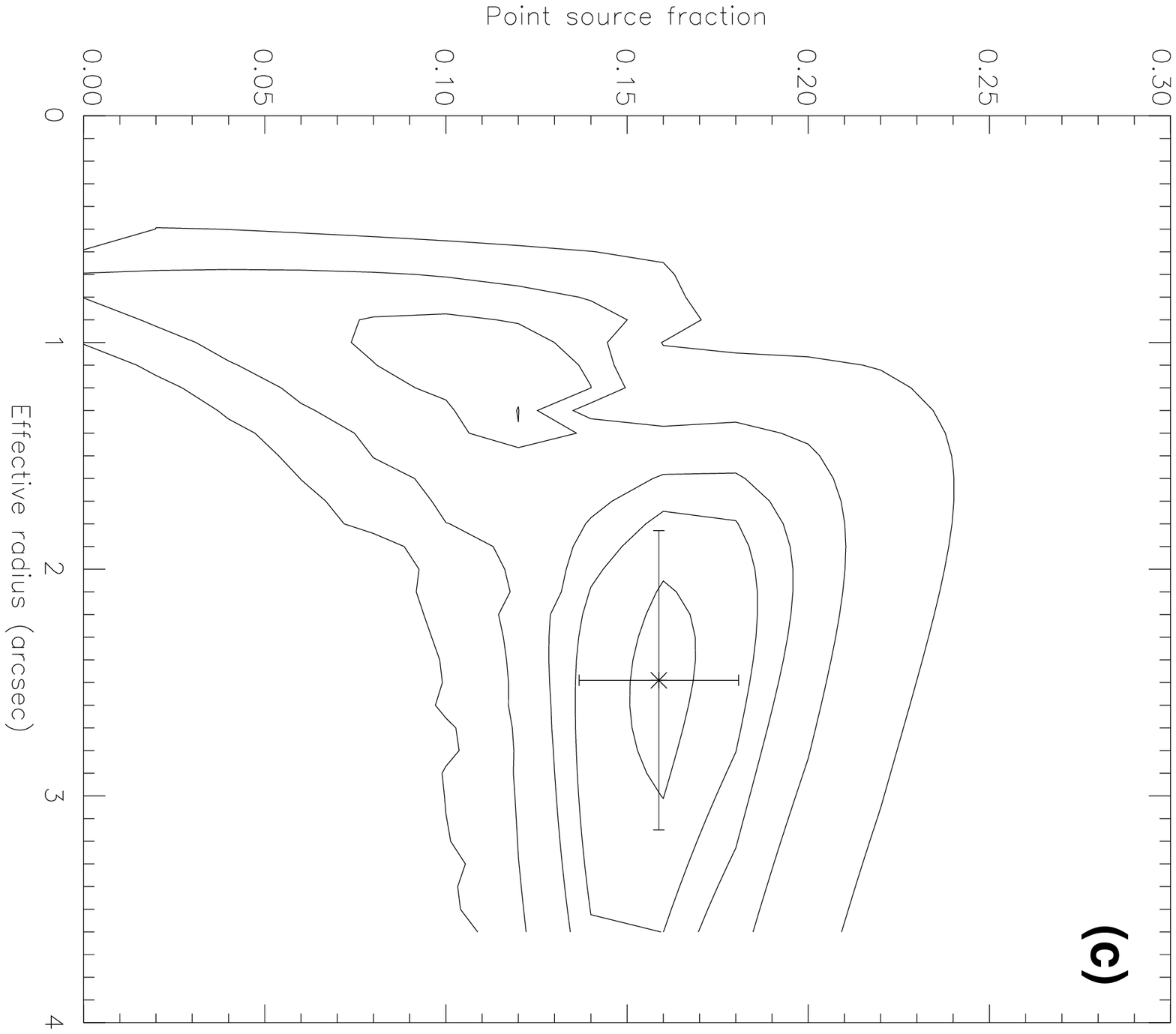}
\end{center}
\caption{Host galaxy morphology fits for PKS1932-46.  The $K-$band
   galaxy image is presented in frame (a), with the residuals after
   best-fit model galaxy subtraction displayed 
   in frame (b). Frame (c) illustrates the  1, 2, 3, 5 and 10-$\sigma$
   contours for the reduced $\chi^2$, plus the best-fit values and
   associated error bars for the
   point source contribution and effective radius.
\label{Fig: morph}}
\end{figure*}

We have modelled the host galaxy morphology within a $32 \times 32\, \rm
arcsec^2$ region using the methods detailed in Inskip
et al. (2005), which can be briefly summarised as
follows.  The point spread function (PSF) for the SOFI field can be
accurately determined 
from unsaturated stars within the mosaic field; their 2-d
profile is extracted, normalised to unit flux, and used to generate an
average PSF profile. In order
to avoid any complications arising from variations in image quality
across the SOFI field of view, we use the nearby M2 star lying to the west of
the galaxy.  Comparison with other stars in the field confirms that
the image quality is generally good, and that this star does not
suffer noticeably from contamination by any nearby faint objects.  
Once a good PSF had been obtained, de Vaucouleurs profile
 galaxy models were convolved with the PSF  and fitted to the surface profile of the galaxy,
using available least squares minimisation IDL
routines\footnote{\textsc{mp2dfunfit.pro}, part of Craig Markwardt's
\textsc{mpfit} non-linear least squares curve fitting package
available via http://astrog.physics.wisc.edu/$\sim$craigm/idl/fitting.html.}.  The free parameters for this
modelling are the galaxy flux,
centroid, effective radius, and fractional nuclear
point source contribution. Galaxy ellipticity was also allowed to
vary, but in the case of this object did not lead to significant
variation in the preferred effective radius, nuclear point source
contribution, or distribution of residual flux.

The modelled data and model-subtracted residuals are displayed in
Fig.~\ref{Fig: morph}, together with a plot of the variation in
reduced $\chi^2$ over the parameter space considered.  The resulting
best-fit parameters give an effective radius of $r_{\rm eff} = 2.49 \pm 0.66$
arcsec (equivalent to $9.18 \pm 2.43$kpc in our assumed cosmological
model), and an unresolved nuclear point source contribution of $15.9 
\pm 2.2$\%.  The residuals (which remain
present at a low level regardless of the model ellipticity) display
some level of asymmetry, with the model slightly underfitting the galaxy
westwards of the nucleus and overfitting it immediately
north and south of the nucleus (see Fig.~\ref{Fig: morph}(b)). These residuals are
likely to reflect the intrinsic asymmetries of the source (as noted in
VM98; e.g. the ``arm'' structure extending eastwards and the N-S
aligned star-forming knots).  Possible causes are contamination of the
galaxy emission by asymmetrically distributed young stellar populations
(which although blue in colour would still be expected to boost the
observed-frame $K-$band flux; see e.g. Inskip et al 2006), or
alternatively, the presence of a dust 
lane lying perpendicular to the radio source axis.  
We also extended our modelling to S\'{e}rsic profiles (S\'{e}rsic
1968; i.e. $r^{1/n}$ rather than simply $r^{1/4}$), so as to
investigate alternative galaxy profiles.   The range of acceptable 
S\'{e}rsic indices for our modelling is $n = 2.75 \pm 1.41$, with other
parameters of $r_{\rm eff} = 5.44 \pm 4.05$ and a point source
contribution of $19 \pm 5$\%.  These relatively large error bars
reflect the fact that there is clearly some level of degeneracy
within the models, and disky elliptical profiles can provide
reasonably good fits to the observed data with similar residuals to
the de Vaucouleurs profile models. 

PKS1932-464 was one of twelve radio galaxies to be studied
by Holt et al (2007), in their detailed modelling of the continuum
properties and stellar populations of the host galaxies.  The galaxy
spectrum is best fit by a combination of a power law and an old stellar
population (aged $\sim12.5$ Gyr); although the addition of a young stellar
population component can also result in good fits to the data, the subsequent
degeneracy between the power law shape and YSP mass/age cannot be
resolved.  Taking their assumed old stellar population and convolving
a GISSEL spectral synthesis (Bruzual \& Charlot 2003) template spectrum with the $K_S$ band
filter profile implies that a galaxy of this age and magnitude
($K_S=15.3$, after removal of point source component; this is close to
the $L\star \sim 15.5$ magnitude expected at this redshift
(e.g. Kochanek et al 2001, Willott et al 2003)) would have a mass
of approximately $2.8 \times 10^{11}M_{\odot}$.  More luminous, younger
stellar populations would of course require less mass; varying the
galaxy age from 5 to 15 Gyr allows for a mass range of $1.7 - 3.2 \times 10^{11}M_{\odot}$.

\begin{figure}
\vspace{1.4 in}
\begin{center}
\includegraphics{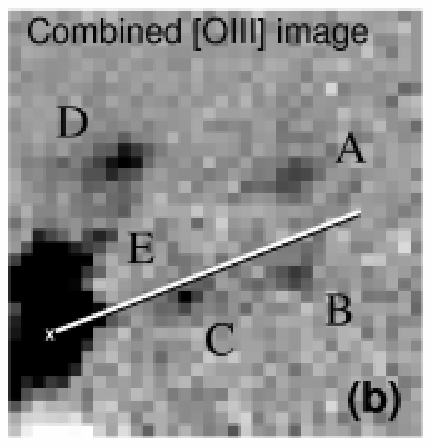}
\includegraphics{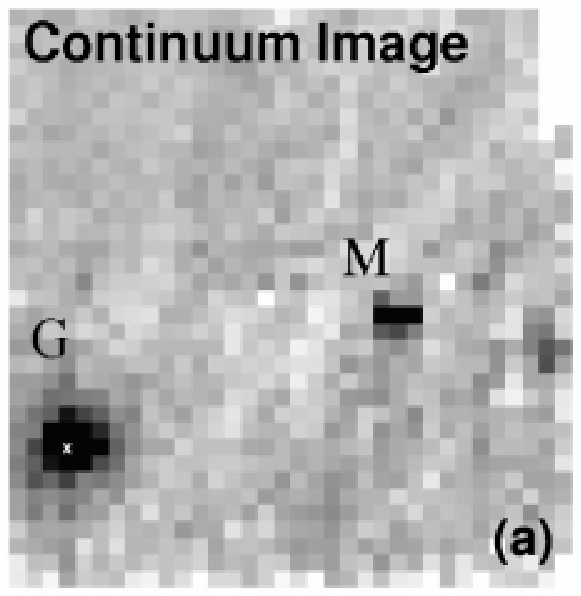}
\end{center}
\caption{(a -- left) Continuum image of PKS1932-46 generated from the
emission either side of the [O\textsc{iii}]5007\AA\ emission line. The
host galaxy and nearby M-star are labeled with the letters G and M
respectively.   (b -- right) Continuum-subtracted
[O\textsc{iii}]5007\AA\ emission  line image generated from summed
data between $\sim$6146\AA\ and $\sim$6174\AA.   Noteworthy emission line features are labelled A-E.  In these figures,
north is to the top and east to the left, the pixel scale is
  $0.67^{\prime\prime}$/pixel (giving a total field of view of $\sim
  23.5 \times 23.5$ square arcsec or $87 \times 87$kpc$^2$), and the radio source
axis (marked with a line illustrating the approximate extent of the
  western lobe) lies at a position angle of $-72^{\circ}$.  The position of the host galaxy centroid is
  marked in both frames with a cross.
\label{Fig: contrast}}
\end{figure}

\subsection{EELR properties}

\begin{figure}
\vspace{4.8 in}
\begin{center}
\includegraphics{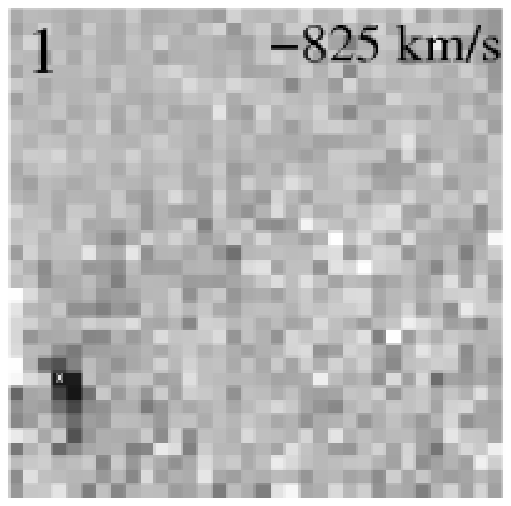}
\includegraphics{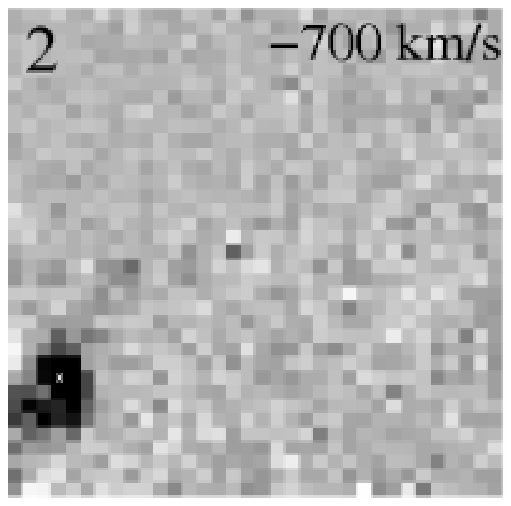}
\includegraphics{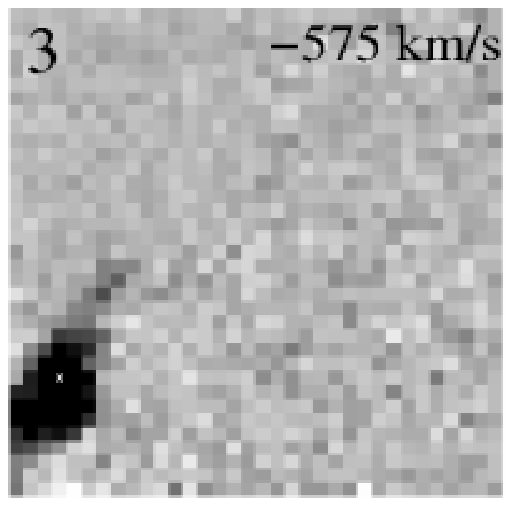}
\includegraphics{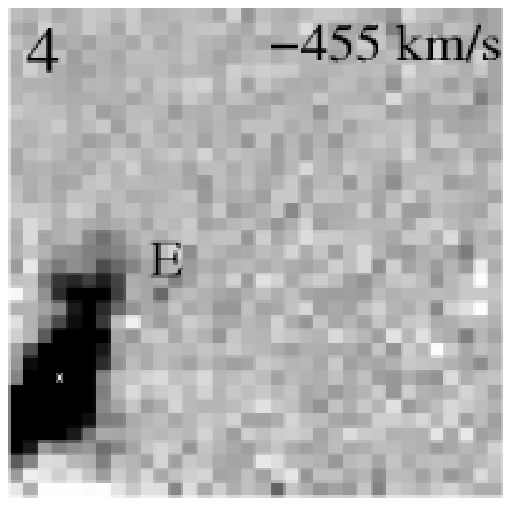}
\includegraphics{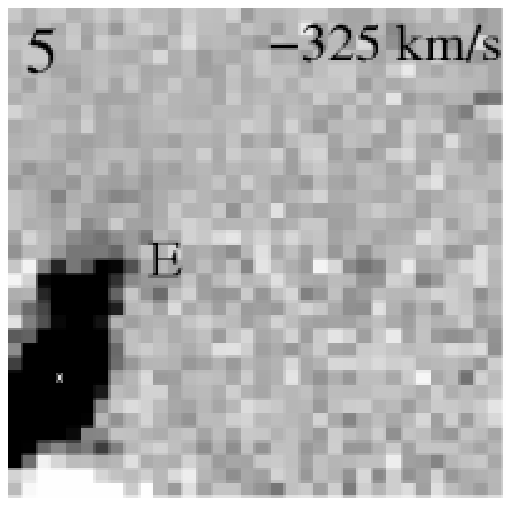}
\includegraphics{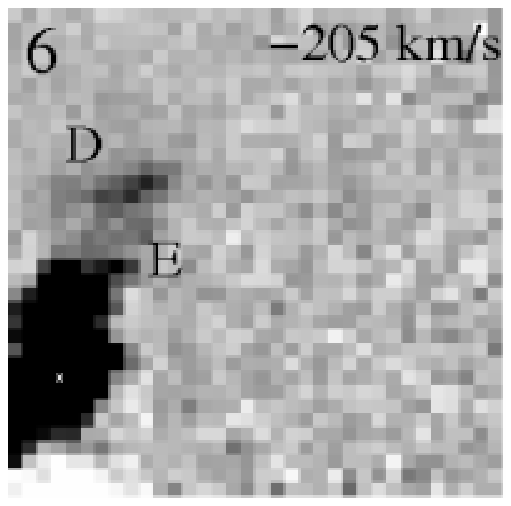}
\includegraphics{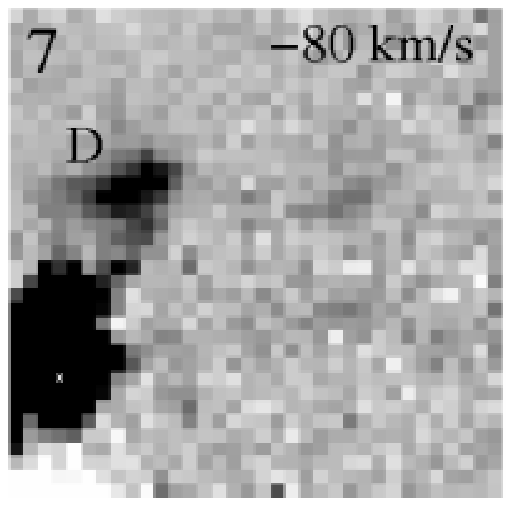}
\includegraphics{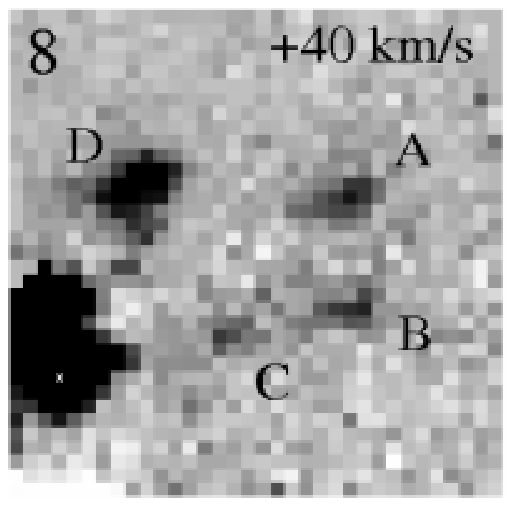}
\includegraphics{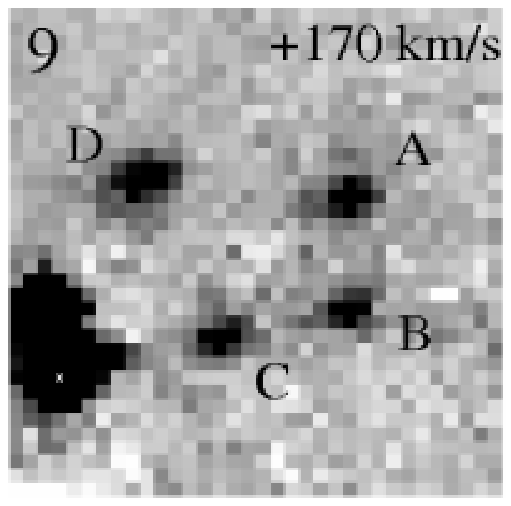}
\includegraphics{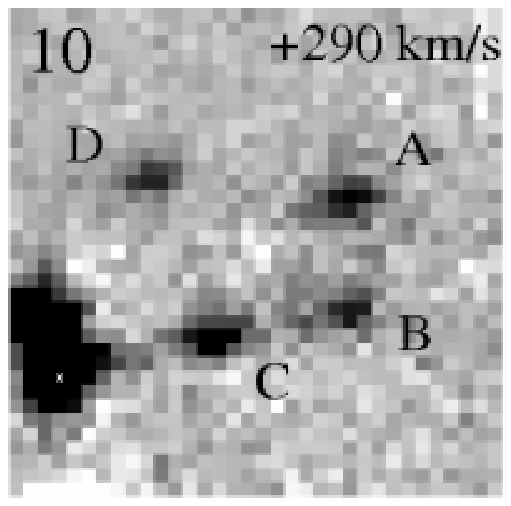}
\includegraphics{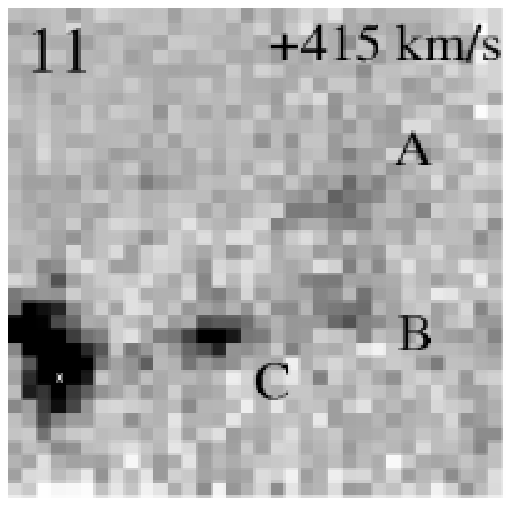}
\includegraphics{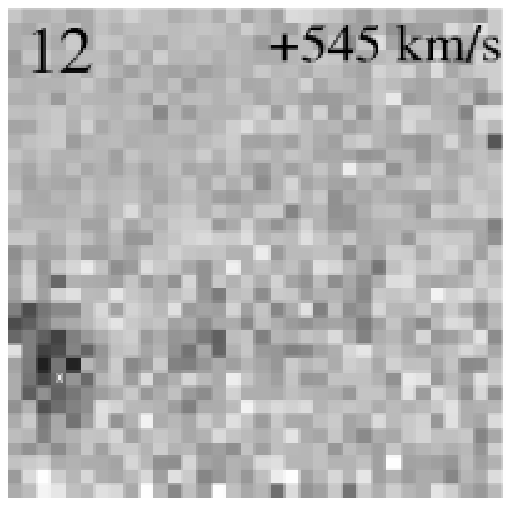}

\end{center}
\caption{IFU flux images of PKS1932-46 and the western EELR generated
from consecutive $\sim$2.5\AA\ spectral bins, from $\sim$6146\AA\ (top left)
to $\sim$6174\AA\ (bottom right).  These wavelengths correspond to
4993--5015\AA\ in the rest-frame of PKS1932-46 (i.e. $z \sim 0.231$), and
cover the [O\textsc{iii}]5007\AA\ line emission.   Labelling of
different features is the same as that used in Fig.~\ref{Fig:
  contrast}, and we also denote the velocity offset of each 
frame relative to the typical redshift of $z=0.231$. The position of the host galaxy centroid is
  marked in all frames with a cross.
\label{Fig: flythru}}
\end{figure}

The EELR lying in and around the western radio lobe of PKS1932-46 is very
complex, and our data reveal a number of feaures that have not been
previously identified. In the following sections we will discuss the
morphology, kinematics and ionisation state of the EELR in turn, making full use
of the versatilty of our IFS data, which is presented in
several different ways in order to best illustrate the different EELR features.

Figures \ref{Fig: contrast} and \ref{Fig: flythru} concentrate on a
small region of the IFS datacube which covers the host galaxy and
western radio lobe.  Fig.~\ref{Fig: contrast} displays a continuum image of
the field and the continuum-subtracted summed [O\textsc{iii}]5007\AA\
emission line image, while Fig.~\ref{Fig: flythru} displays the
continuum-subtracted flux stepped in equal $\sim 2.5$\AA\
spectral bins (equivalent to $\sim$110km\,s$^{-1}$) over the wavelength range 6146-6174\AA\ (i.e. covering 
the [O\textsc{iii}]5007\AA\ emission line).  This approach allows the
continuum and
line emitting regions to be considered separately, and provides a
clear image of the varied velocities of the line-emitting material.

In order to study the EELR properties in greater detail, we have developed IDL routines
to model the line emission on a fibre-by-fibre basis.  For each fibre
spectrum, we locate a selected emission line and fit it with one or more Gaussian
components plus a continuum, extracting the resulting modelled emission-line flux, FWHM and velocity offset
relative to the systemic redshift of PKS1932-46. The resulting data are
displayed in Fig.~\ref{Fig: 2}, and provide a more global overview of
the kinematics of the system.   Fig.~\ref{Fig: 2} also includes data
from the companion galaxy lying to the northeast of PKS1932-46.

\subsubsection{EELR morphology}

\begin{figure}
\vspace{6.05 in}
\begin{center}
\includegraphics{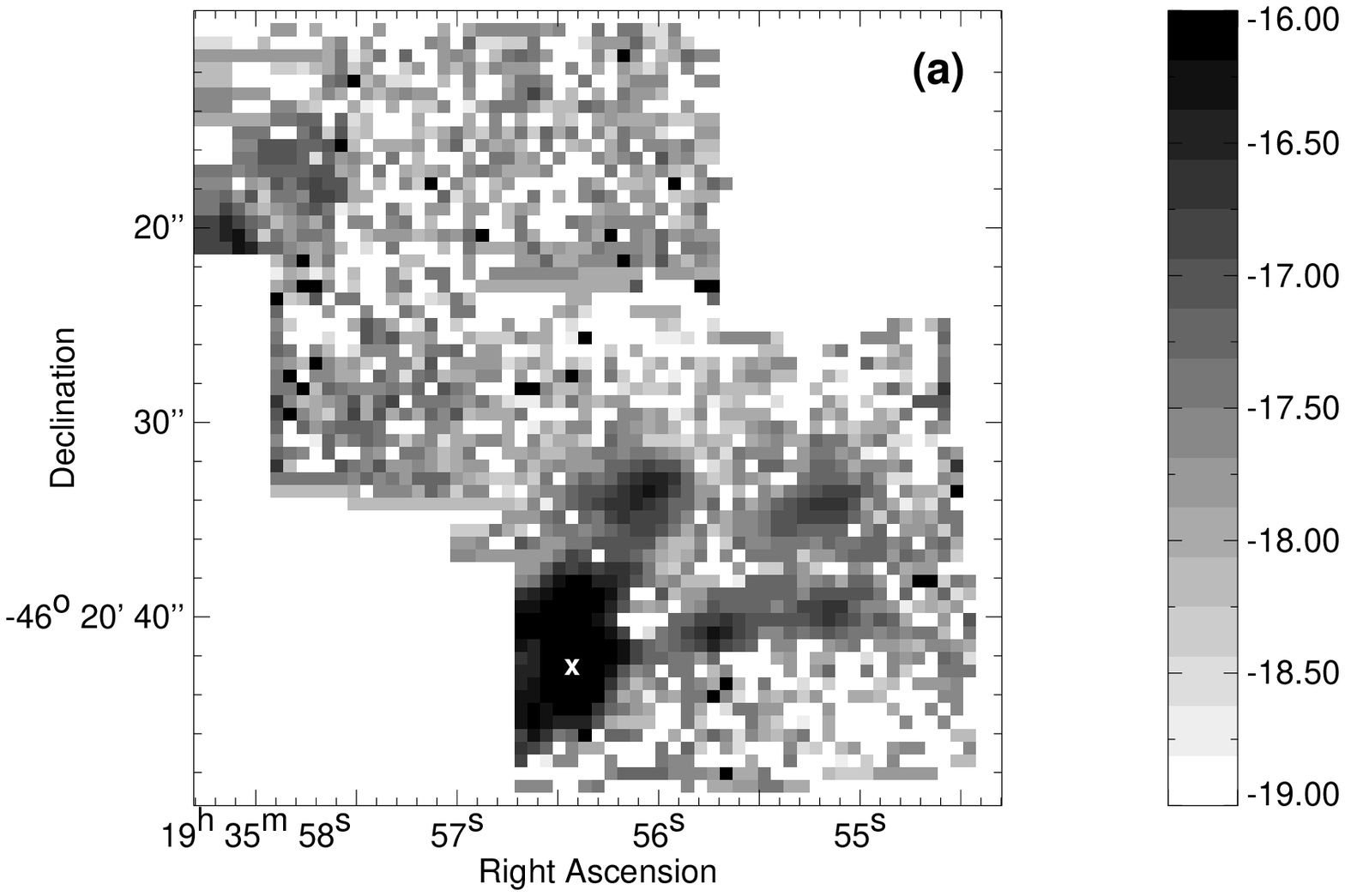}
\includegraphics{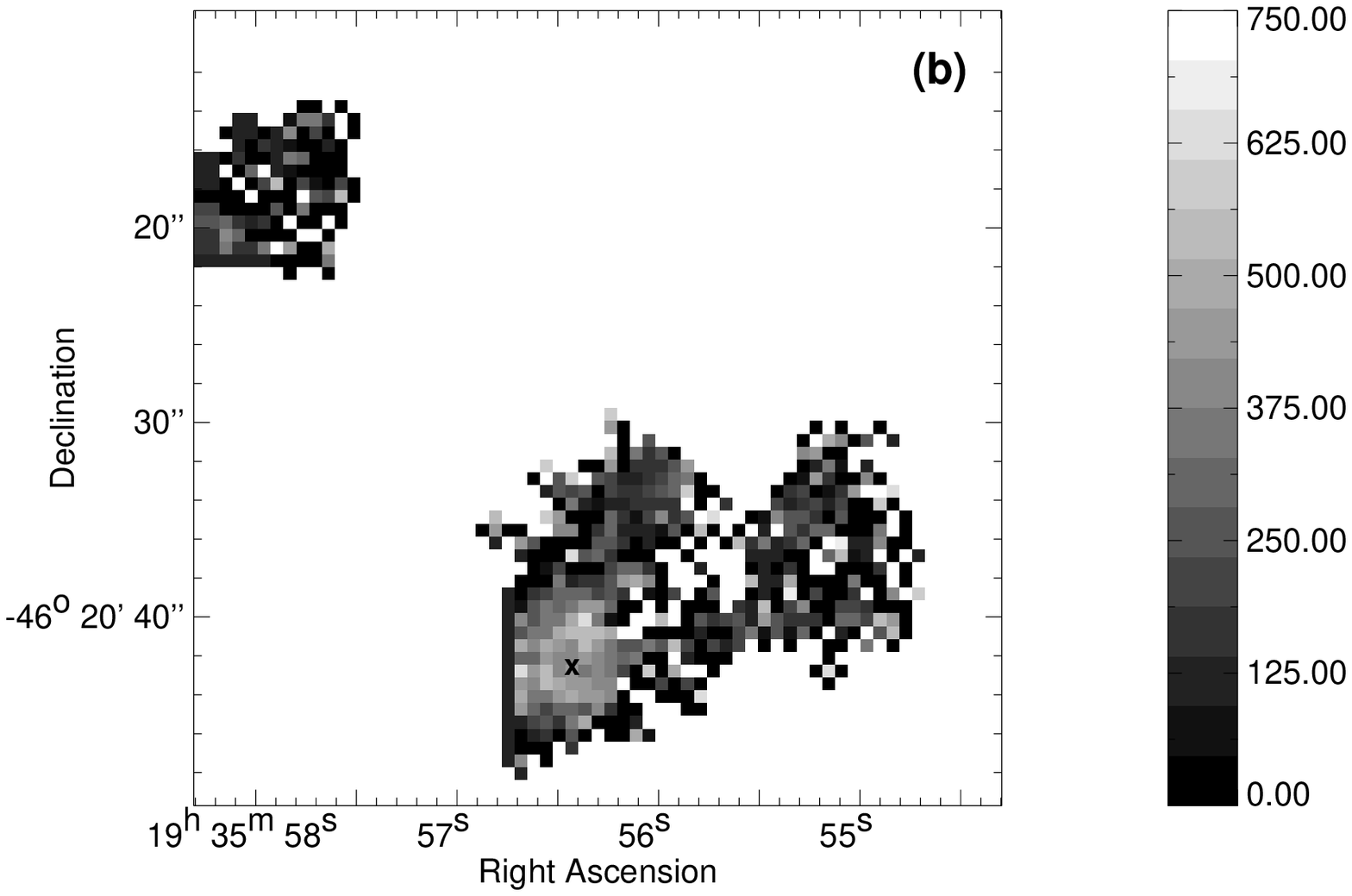}
\includegraphics{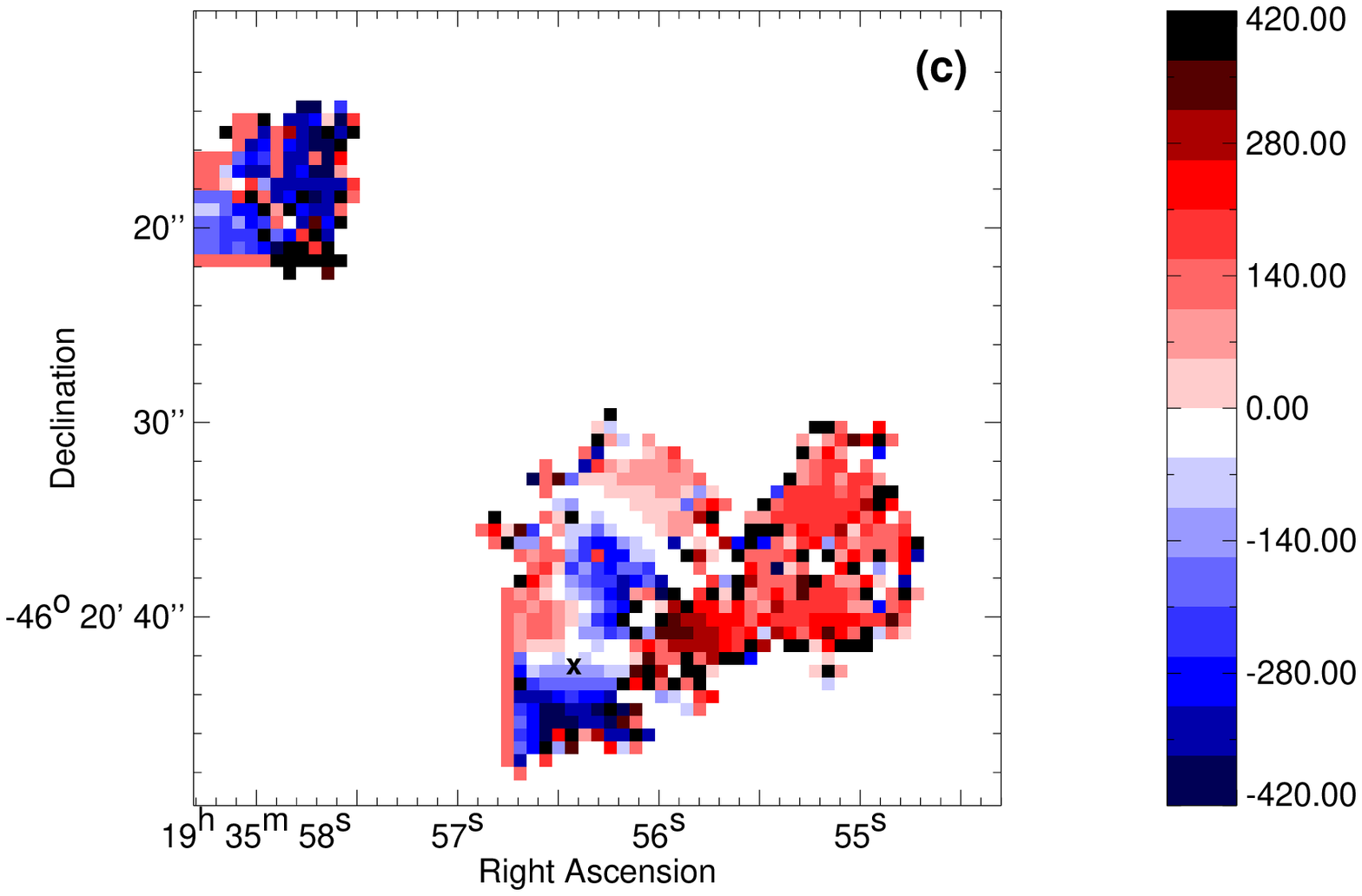}
\end{center}
\caption{From top: (a) Extracted [O\textsc{iii}]5007\AA\ line
  flux; (b) extracted [O\textsc{iii}]5007\AA\ FWHM line
  width (corrected for instrumental broadening); (c) velocity shift of
  [O\textsc{iii}]5007\AA\ relative to $z=0.231$ (the approximate rest frame
  of the host galaxy). The position of the host galaxy centroid is
  marked in all frames with a cross.  The grey-scale bars
  illustrate either log(flux) in units of $\rm{erg\,s^{-1}} cm^{-2}$ or
  velocities in \kms.  For the plots of FWHM and velocity
  offset, we have masked the fibres with an [O\textsc{iii}] emission line flux below $1
  \times10^{-18} \rm{erg\,s^{-1}} cm^{-2}$.
\label{Fig: 2}}
\end{figure}

As Fig.~\ref{Fig: contrast} and Fig.~\ref{Fig: 2} illustrate, the EELR
surrounding PKS1932-46 extends throughout the western radio lobe, and
contains a number of discrete, spatially resolved features in addition
to the powerful line emission centred on the host galaxy itself.
However, the full complexity of the EELR is only revealed once we can
disentangle the different velocity components in the emitting
material.  Figure \ref{Fig: flythru} illustrates twelve consecutive
continuum-subtracted slices
of the IFS datacube covering the [O\textsc{iii}]5007\AA\ emission
line. The western radio lobe contains a series of resolved clumpy 
features (labelled A-C in Figs.~\ref{Fig: contrast}b and
\ref{Fig: flythru}) which produce [O\textsc{iii}]5007\AA\ emission at the
longest wavelengths.  While we cannot pin
down precise locations of the emission features relative to the radio
source cocoon along the line of sight, the observed features in the
western lobe all have very similar redshifts, and lie at a variety of
distances between the host galaxy and the radio hotspot.  

The bright knots to the north of the host galaxy (labelled D \& E in
Figs.~\ref{Fig: contrast} and \ref{Fig: flythru}) lie at a considerable
angular distance from the radio source axis:  knot D lies $\sim 10^{\prime\prime}$/35kpc away
from the AGN along a PA of $\sim -30^{\circ}$,
and has been detected in previous emission line imaging
(VM98).  As Fig.~\ref{Fig: flythru} illustrates, it can be immediately discerned that these knots lie
at different redshifts to the other EELR knots in the western lobe,
and that there appears to be a velocity gradient between knots D and E.  

The central regions of the EELR display the most intense line
emission, at several different velocities.  In addition to 
the ``arm'' structrure (Fig.~\ref{Fig: flythru} frame 10)  lying to the
east of the host galaxy which was noted in previous observations
(VM98), we also observe blue-shifted emission (including knot E;
Fig.~\ref{Fig: flythru} frames 5 \& 6) 
extending approximately north--south roughly {\it perpendicular} to the
radio source axis.  Part of this blueshifted region was detected in
the spectroscopic observations of VM05.

\subsubsection{EELR kinematics}

Fig.~\ref{Fig: 2} presents a quantified view of the
[O\textsc{iii}]5007\AA\ emission line data (flux, line 
width and velocity) for PKS1932-46 obtained via single Gaussian plus continuum fitting.
The material lying to the west of the host galaxy displays the
narrowest line widths, typically with measured FWHM of $\sim200$\kms
(all line widths are corrected for instrumental broadening). The largest line widths (up
to $\sim300-750$\kms) are observed towards the centre of the host
galaxy, consistent with the line widths observed for [O\textsc{iii}]
observed in VM98 and VM05. However, given the complex kinematics observed in these
regions, single Gaussian fits to the line profile do not necessarily 
provide the most reliable measure of the true emission line FWHM. 

In terms of the relative velocities of the different emission
features, the knots in the western lobe are distinctly redshifted relative
to the systemic velocity of the system, as can also be observed in
Fig.~\ref{Fig: flythru}. The line emission from the
northern knot (labelled D in Figs.~\ref{Fig: contrast} and \ref{Fig:
  flythru}) is quite different; as the velocity map illustrates, there
appears to be a fairly smooth velocity 
gradient of up to $\sim400$km\,s$^{-1}$ across this feature, with the
material lying closest to the host galaxy emitting at shorter
wavelengths than the more distant material.  Similar variations are
observed across the central regions of the EELR, as expected from the
superposition of the structures observed in Fig.~\ref{Fig: flythru}. 

\begin{figure}
\vspace{2.1 in}
\begin{center}
\includegraphics{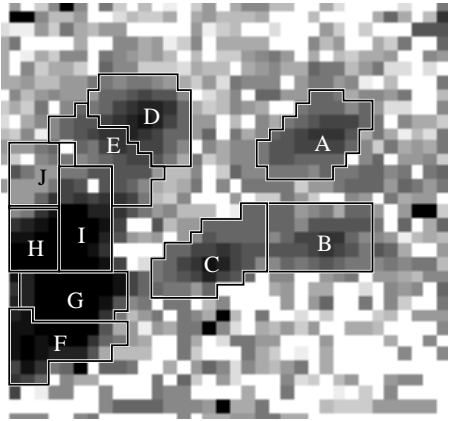}
\end{center}
\caption{[O\textsc{iii}] flux image illustrating the ten multi-fibre
  regions extracted as part of our analysis of weaker emission lines.
 In the outer parts of the EELR, each region reflects an individual
  emission feature.  In the central regions, we have selected fibres
  according to similar kinematic properties.  The field of view of
  this image is approximately $23.5 \times 23.5$ arcsec squared.
\label{Fig: region_ID}}
\end{figure}

In order to analyse our IFU spectra at a higher signal--to--noise
level and address the issue of superposed velocity structures,
we have extracted and combined the spectra of fibres covering specific
emission regions, and those which display similar kinematic
properties.  The resulting regions, labelled A to J, are displayed in
Fig.~\ref{Fig: region_ID}.  It should be noted that regions D and E on
Fig.~\ref{Fig: region_ID} cover feature D in figures 4 and 5,
while region I (and to a lesser extent, J) on Fig.~\ref{Fig:
  region_ID} corresponds most closely to feature E in figures 4 and 5.
We have re-analysed the
[O\textsc{iii}]5007\AA\ line widths and velocity shifts, fitting up to
three Gaussians to the combined emission line data.  The results of
our kinematic analysis are consistent between different lines, the
earlier spectroscopy of VM98, and with
the single-fibre data displayed in Fig.~\ref{Fig: 2}; we 
tabulate the results for [OIII]4959,5007 in Table 3.  

\begin{figure*}
\vspace{2.1 in}
\begin{center}
\includegraphics{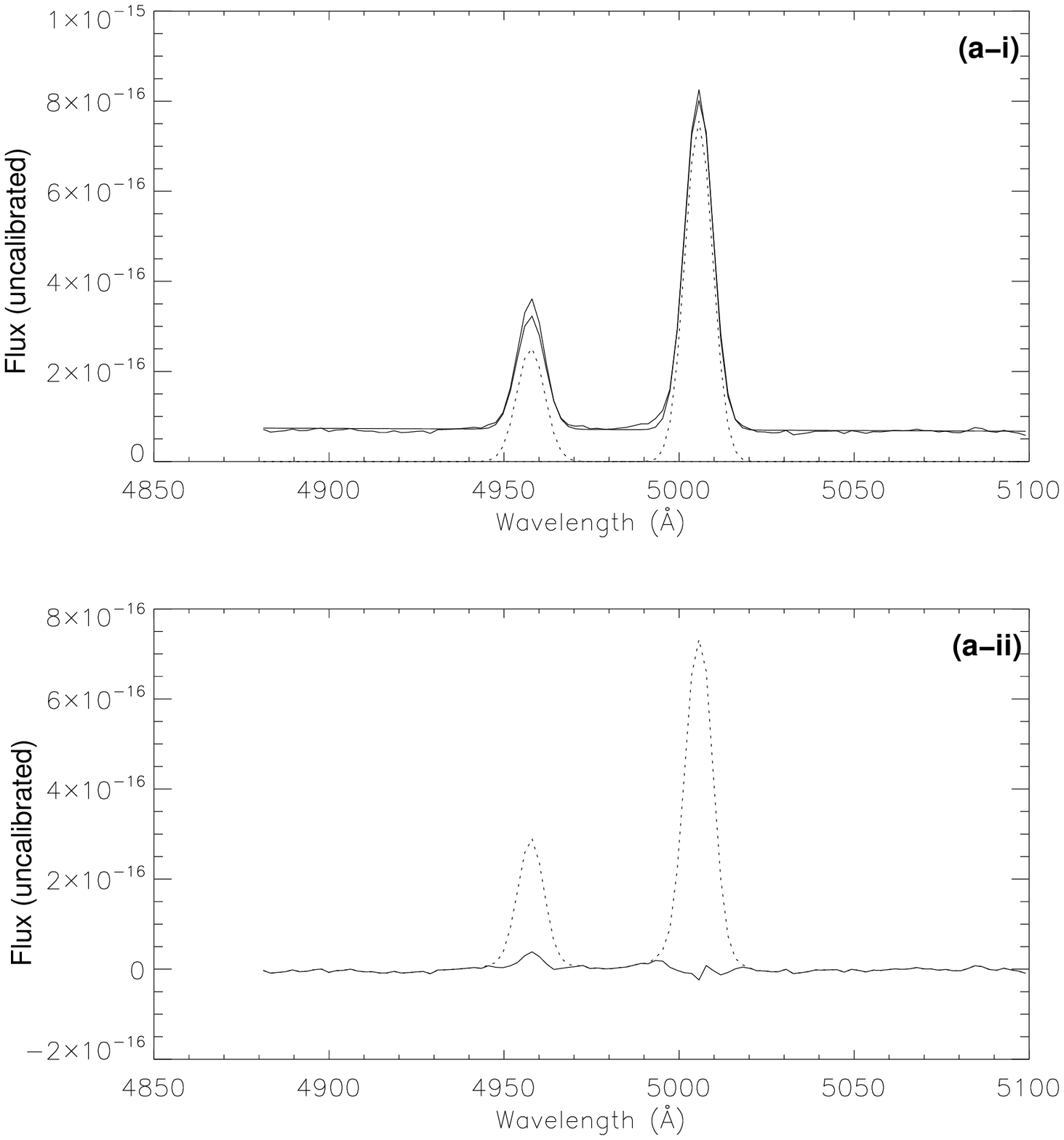}
\includegraphics{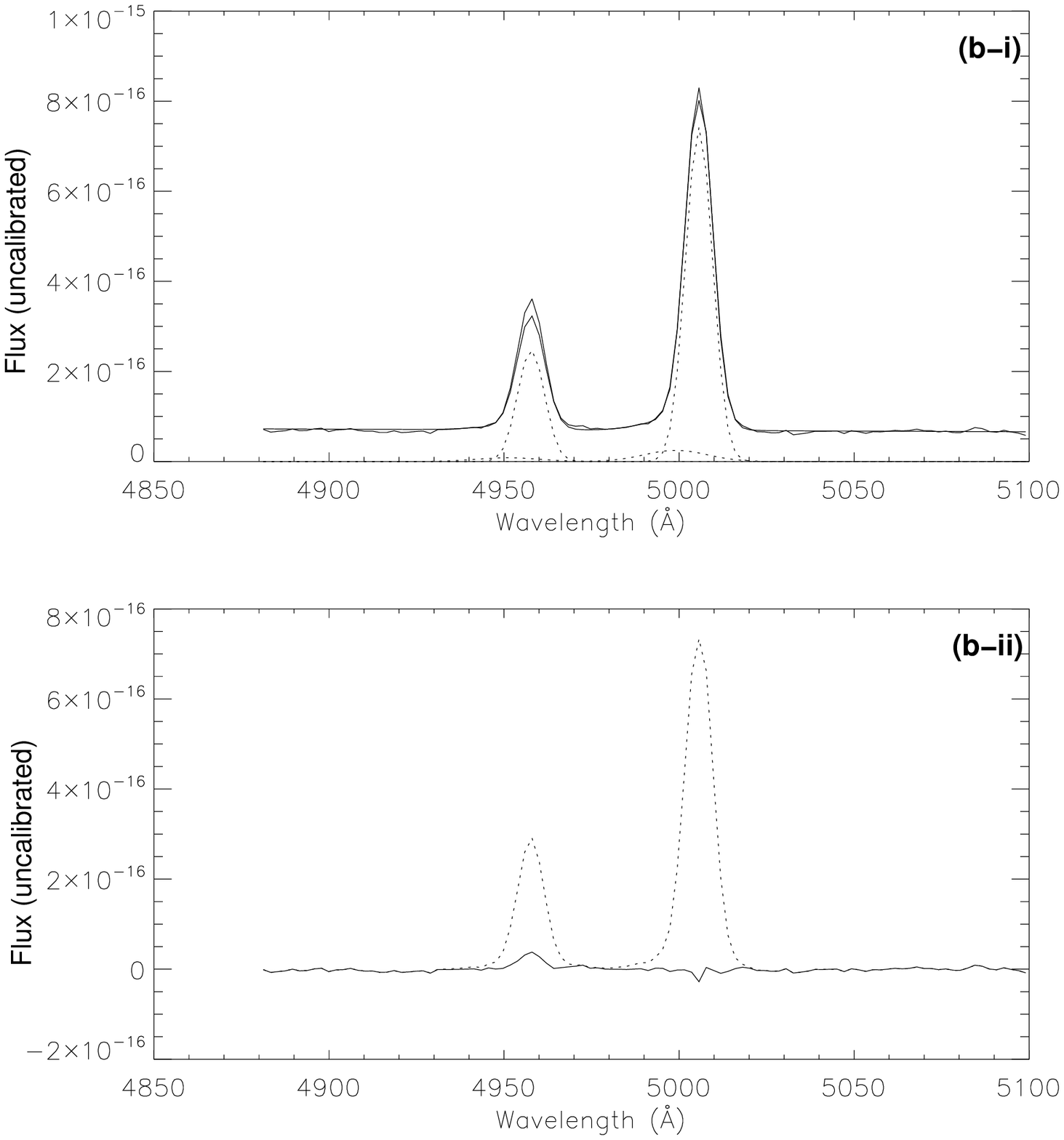}
\includegraphics{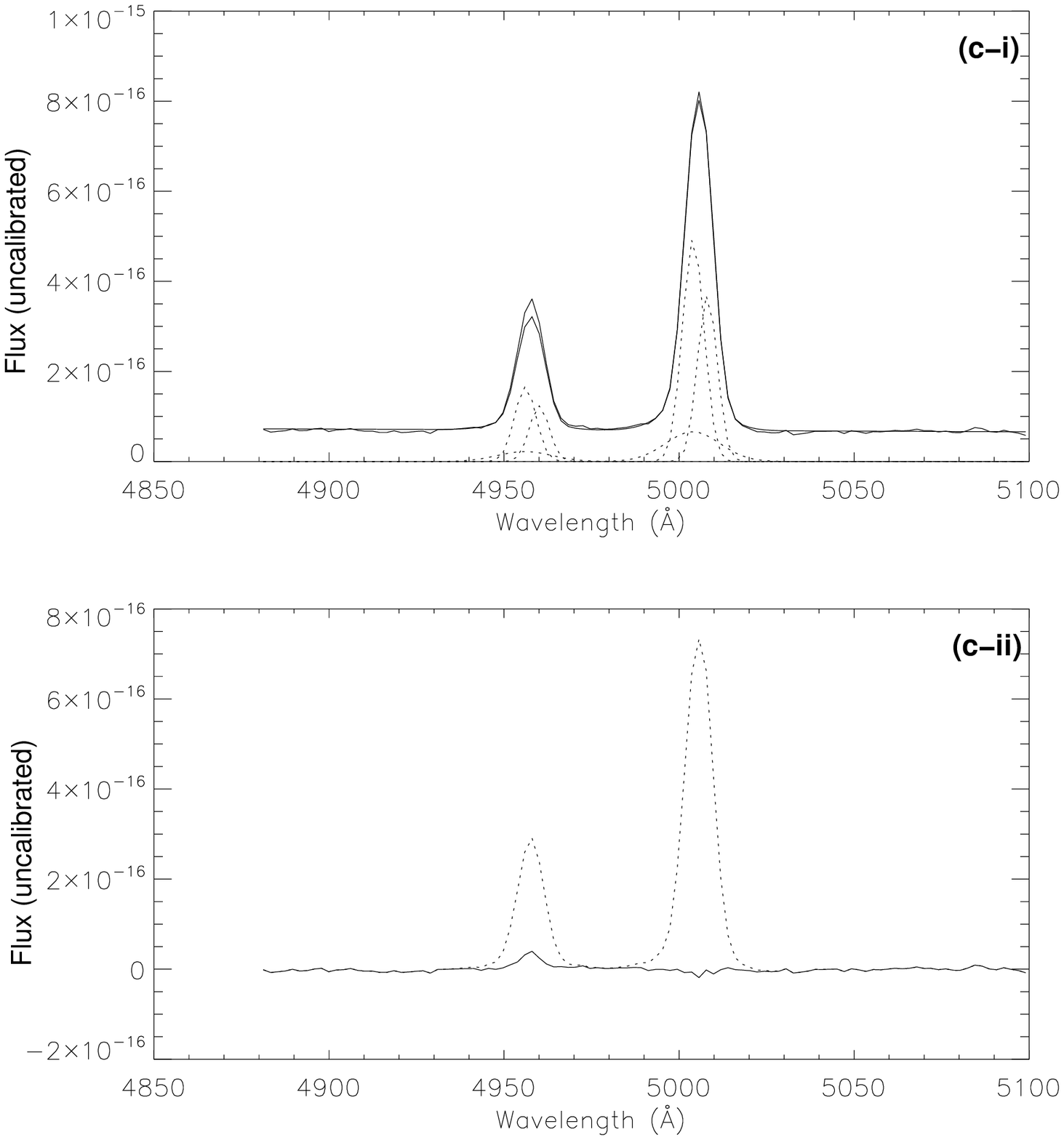}
\end{center}
\caption{Comparison between fits of the O[\textsc{iii}]5007\AA\ emission from region G using
  different numbers of Gaussian components: a single Gaussian (a - left),
  two Gaussians (b - centre) and three Gaussians
  (c - right).  In the top row (a-i, b-i, c-i), we display the 
  observational data and the model fit (solid lines) together with the
  individual Gaussian components (dotted line).  On the bottom row
  (a-ii, b-ii, c-ii), we
  display the observational data (dotted line) and the residuals after
  model subtraction (solid line).  
\label{Fig: Kinematics}}
\end{figure*}

In general,  multiple Gaussian line profiles do not provide a better
fit to the data than single Gaussian models. Most of the extended
emission line regions appear to have relatively quiescent kinematics,
i.e. small line widths and velocity shifts.   The exceptions to this are
regions G and I (see Fig.~\ref{Fig: Kinematics}), where models combining
multiple emission line components are more plausible.  In these
regions closer to the galaxy nucleus, the gas kinematics are more complex, and there are at least
two clear velocity components (see Fig.~\ref{Fig: flythru}): (i) the
blue-shifted emission which lies in 
a north-south direction across regions I, G, F, and (ii) the ``arm''
feature (VM98) projecting eastwards from regions G \& I at roughly the
systemic redshift.   Unfortunately, as is often
the case for emission line fitting, there is some degeneracy between
fits with two or more narrow components, and those combining both
narrow and broader emission features.   In addition to fits combining
two narrow components, the preferred models for both the two- and
three-component fits of region G also include a broader emission
component (FWHM $\sim 1000-1200$\kms; also observed in VM05), which is blueshifted relative to the
systemic redshift by $\sim$ 200-500\kms.  

\begin{table*}
\caption{Kinematics of the multi-fibre regions illustrated in
  Fig.~\ref{Fig: region_ID}. Velocity shifts are relative to
  the typical value of $z=0.231$.  Except for regions G and I
  (where 2 and 3 Gaussian fits are also presented), only a
  single Gaussian component is required for each region.  }
\begin{tabular}{lccc}
Region & Line Width (FWHM)  & Velocity Shift & Reduced $\chi^2$\\
& ($km\,s^{-1}$) & ($km\,s^{-1}$) &\\\hline
A& $189 \pm 20$& $+181 \pm 34$ & 1.40\\
B& $198 \pm 130$& $+203 \pm 47$ & 1.50\\
C& $215 \pm 21$& $+261 \pm 35$ & 0.90\\
D& $229 \pm 9$& $+51 \pm 34$ & 1.27\\
E& $153 \pm 22$& $-65 \pm 34$ & 1.08\\
F& $345 \pm 15$& $-319 \pm 35$ & 1.09\\
G (single)& $440 \pm 2$& $-82 \pm 34$ & 2.10\\
G (double)& $442 \pm 2, 1208 \pm 118$& $-77 \pm 34$, $-480 \pm 66$ & 2.02\\
G (triple)& $179 \pm 7, 302 \pm 3, 1067 \pm 43$& $+87 \pm 34$, $-151
\pm 34$, $-189 \pm 40$ & 1.67 \\
H & $322 \pm 15$& $+87 \pm 34$ & 1.07\\
I (single)& $417 \pm 9$& $-134 \pm 34$ & 1.22\\
I (double)& $280 \pm 15, 512 \pm 20$& $-190 \pm 34$, $-37 \pm 35$ & 1.00\\
J& $148 \pm 178$& $+95 \pm 49$ & 1.10\\
\end{tabular}
\end{table*}

\begin{figure*}
\vspace{4.6 in}
\begin{center}
\includegraphics{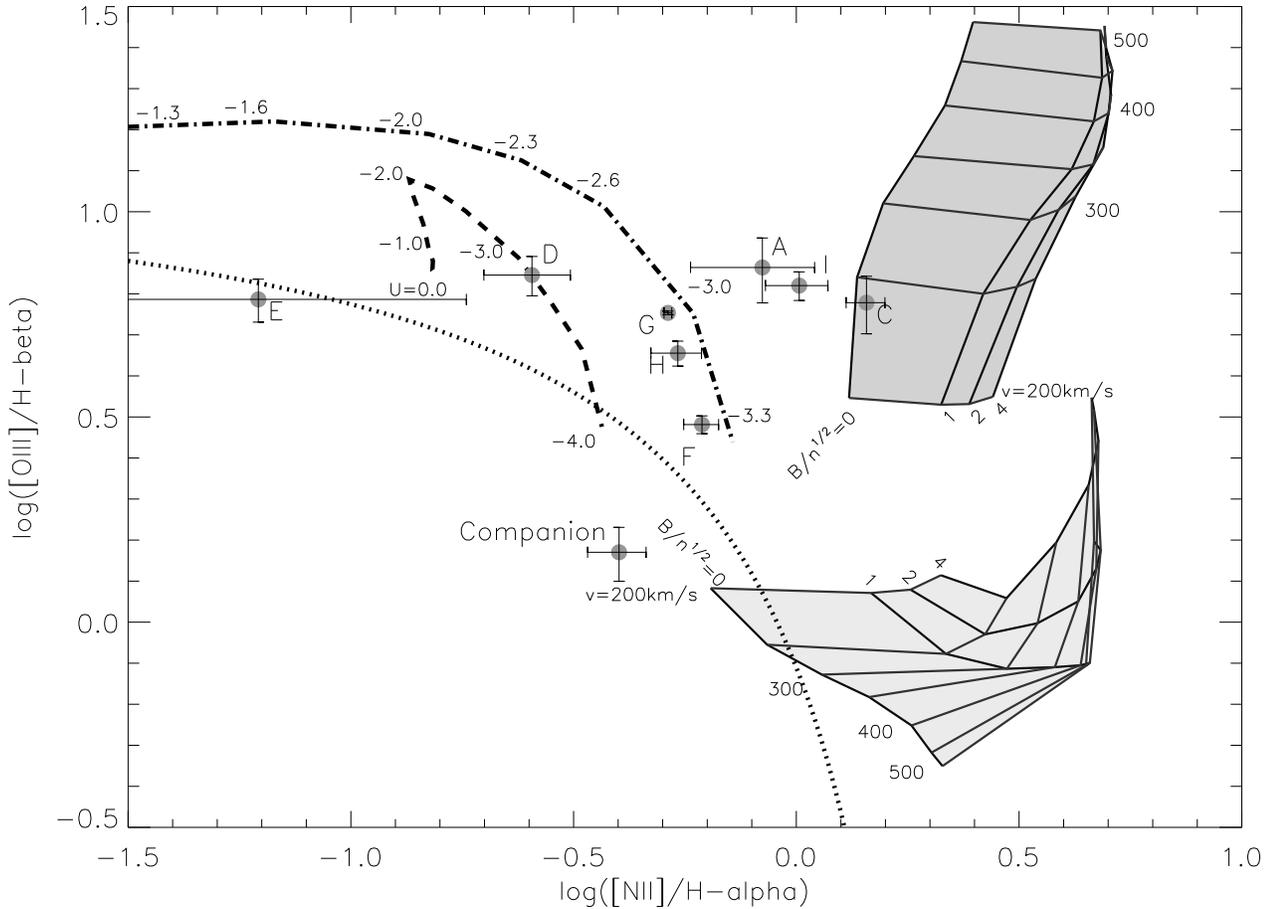} 
\end{center}
\caption{[O\textsc{iii}]/H$\beta$ vs. [N\textsc{ii}]/H$\alpha$
  emission line diagnostic diagram.  The dotted
  line is the maximum starburst track of Kewley et al (2001).  We also
  display the predictions of AGN
  photoionisation models (Groves, Dopita \& Sutherland
  2004) both excluding and including the effects of
  dust (dot-dashed and dashed lines respectively) for an electron density of $n_e=100\rm cm^{-3}$ and
  a range of ionisation parameters ($0.0 > $log$_{10}U > -4.0$). The
  grey regions represent the predictions of the shock models 
  of Dopita \& Sutherland (1996), with (dark grey) and without (light
  grey) a photoionising precursor region.  The data points are for the
  regions illustrated in Fig.~\ref{Fig: region_ID}; regions B and J are absent
  due to large uncertainties, particularly on the [N\textsc{ii}] emission line strength.
\label{Fig: BPT}}
\end{figure*}

\subsubsection{EELR ionisation}

Emission line ratio diagnostic diagrams provide a useful means of
linking the observed line strengths to different ionisation
mechanisms and ionising sources.  The [O\textsc{iii}]/H$\beta$ versus
[N\textsc{ii}]/H$\alpha$ line ratio diagram (Baldwin, Phillips \&
Terlevich 1981) is one of the earliest and most popular diagrams of
this sort, and can clearly distinguish between photoionised line emission from star
forming galaxies, LINER galaxies and Seyferts/AGN.  However, it should
be noted that the use of a single line ratio diagram is 
inherently dangerous; the position of observational data relative to
the predictions of different models does not always remain consistent
between one diagram and another involving different emission line
pairs (e.g. Inskip et al 2002b).  These caveats aside, this diagram can
still be used to shed light on the nature of the varied EELR
components in our data.

While our data at observed-frame wavelengths shorter than 7200\AA\, (including the
[O\textsc{iii}]4959,5007\AA, [O\textsc{iii}]4363\AA\ and H$\beta$
emission lines) are very clean, at longer wavelengths the data can be
greatly affected by CCD fringing and the presence of bright sky lines
(as was explained in some depth in section 2.1), and thus are subject
to larger uncertainties in their relative fluxes. 
In the case of our second IFU data cube (which subtracts the flux from one pointing
from the alternative pointing) the fibres for several regions of
interest (i.e. regions D, E and J) contain object flux in {\it both}
pointings. We therefore use the first
data cube to determine the [N\textsc{ii}]/H$\alpha$ line ratios in these
regions.  As this alternative data cube is imperfectly corrected for
fringing, the errors on this line ratio are noticeably larger for
these regions.

Fig.~\ref{Fig: BPT} displays the  [O\textsc{iii}]5007/H$\beta$
vs. [N\textsc{ii}]6583/H$\alpha$ line ratio diagram for our IFS data.
The observed line ratios are  generally consistent with AGN
photoionisation for the regions F, G and H.  The western knots (A, C)
and regions I and J exist in a noticeably different ionisation state, between the
predictions of AGN photoionisation and the shock models with a
photoionising precursor. Some interaction with the radio source may be
plausible in the case of region C, although the quiescent kinematics
in this region would argue against the presence of significant direct
shock ionisation.  Although the more varied kinematics observed in the central
regions (F, G, H, I) might be suggestive of
shocks, the line ratios for these regions do not place them
significantly closer to the predictions of shock models except in the
case of region I.  The
[N\textsc{ii}]/H$\alpha$ ratios of the blueshifted 
gas components (D, E) are also relatively low.  Region D lies at a
significant  offset from the predictions
of simple AGN photoionisation (dot-dashed track) which are compatible with regions G, H
and F, but is well explained by AGN photoionisation including the
presence of dust (dashed track).  Region E is quite close
to the maximum starburst track of Kewley et al (2001) on
Fig.~\ref{Fig: BPT}, and at a significant offset from the predictions
of simple AGN photoionisation. Based on the emission
line imaging of VM98, these regions are noticeably bright in
[O\textsc{ii}]3727\AA\ emission, also indicating a fairly low
ionisation state.  It is most likely that the emission from these
regions can be interpreted as being due to stellar photoionisation.
This is consistent with the spectroscopic observations of VM05 (obtained using a slit PA of
$-5^{\circ}$, i.e. coincident with region E and the eastern side of
region D), which identified a number of star forming regions in
this part of the EELR.

\begin{figure}
\vspace{2.45 in}
\begin{center}
\includegraphics{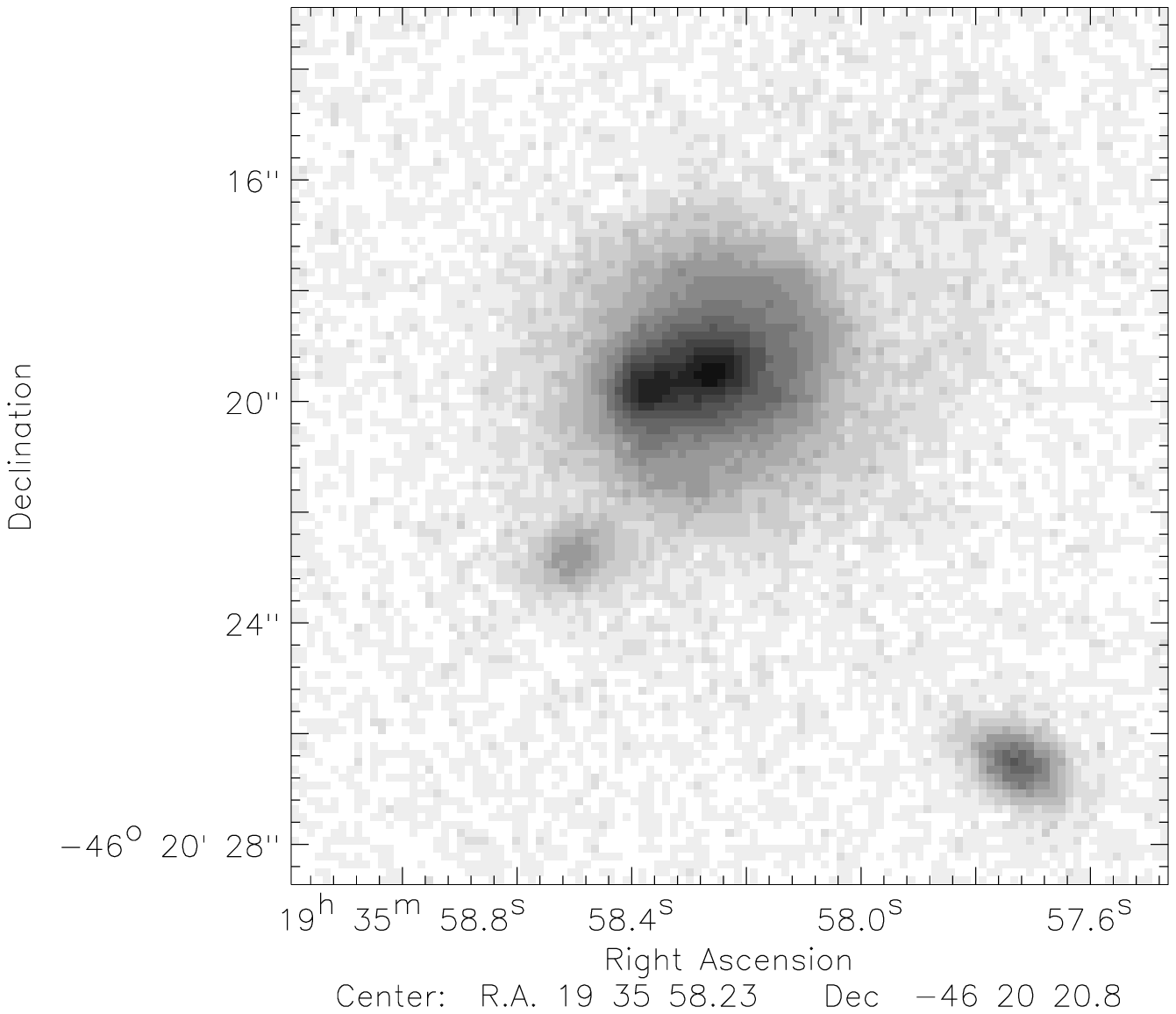}
\end{center}
\caption{$K_S-$band image of the companion galaxy, revealing several very
  interesting morphological features: a double nucleus, tidal tail
  (extending northwards from the west ofthe galaxy; see also Fig~\ref{Fig: kband}),
  and several spatially resolved objects which may possibly be dwarf satellites.
\label{Fig: evilchicken}}
\end{figure}


\subsection{Spitzer MIPS imaging, and the z=0.229 companion galaxy}

It has been speculated in the past (VM98) that the galaxy lying to the
north of PKS1932-46 could lie at a similar redshift.  Previous
imaging of this source suggested the possible presence of spiral arms,
on the basis of the bright emission feature to the south-east, and the 
low surface brightness structure extending to the north-west.  Our
$K_S-$band imaging data (Figs.~\ref{Fig: kband} and \ref{Fig:
  evilchicken}) represents a 
significant improvement in resolution and signal-to-noise, and clearly
shows that this galaxy has a far more complicated morphology.
Figure~\ref{Fig: evilchicken} displays the $K_S-$band image of this
galaxy after adjusting the contrast levels.  The north-west feature
has the appearance of a tidal tail rather than a spiral arm, and the
south-east emission feature is revealed as a separate object.  More
interestingly, the galaxy itself has a very distinctive double
nucleus; this galaxy is almost certainly a merging system of some
description. 

\begin{figure}
\vspace{3.6 in}
\begin{center}
\includegraphics{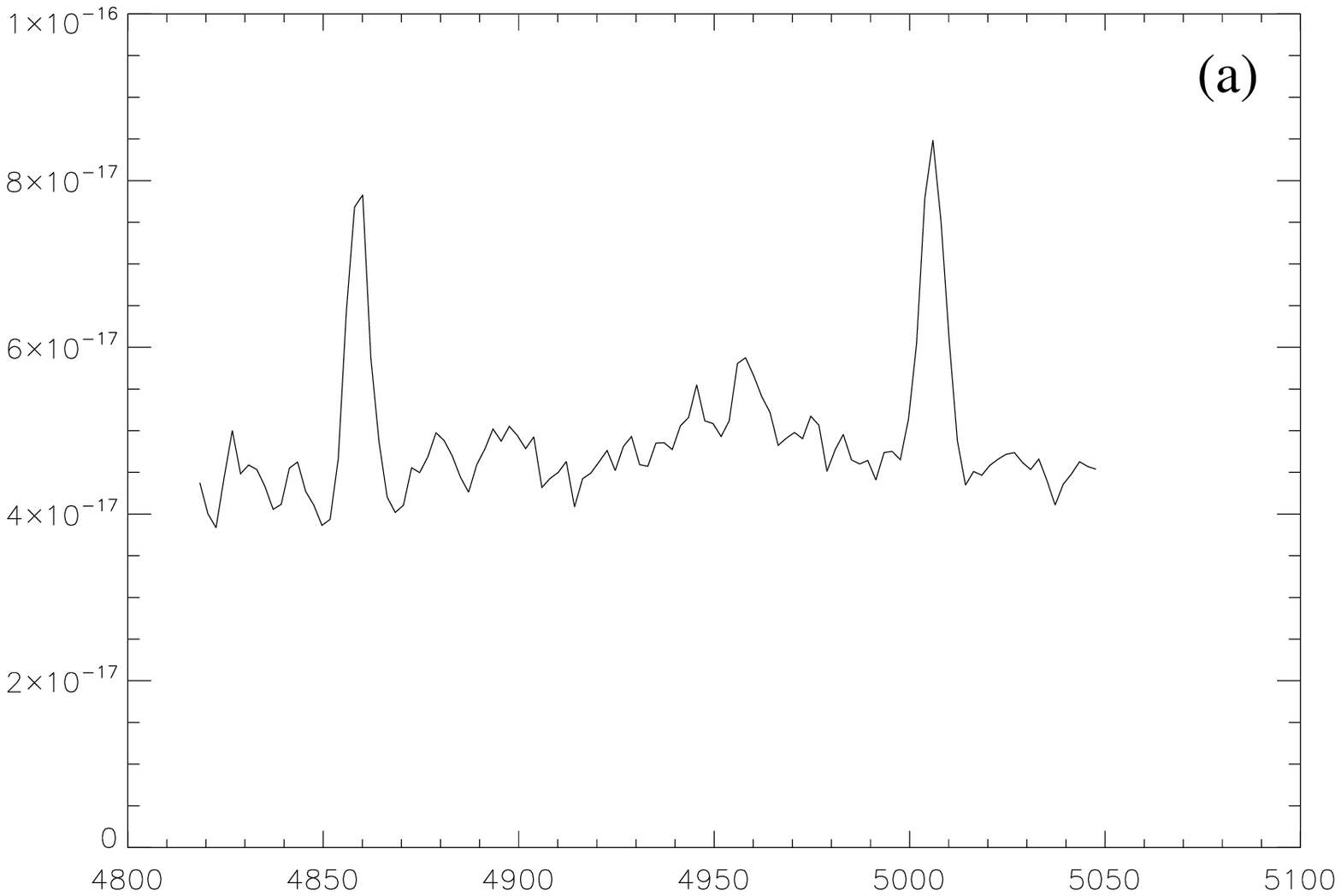}

\includegraphics{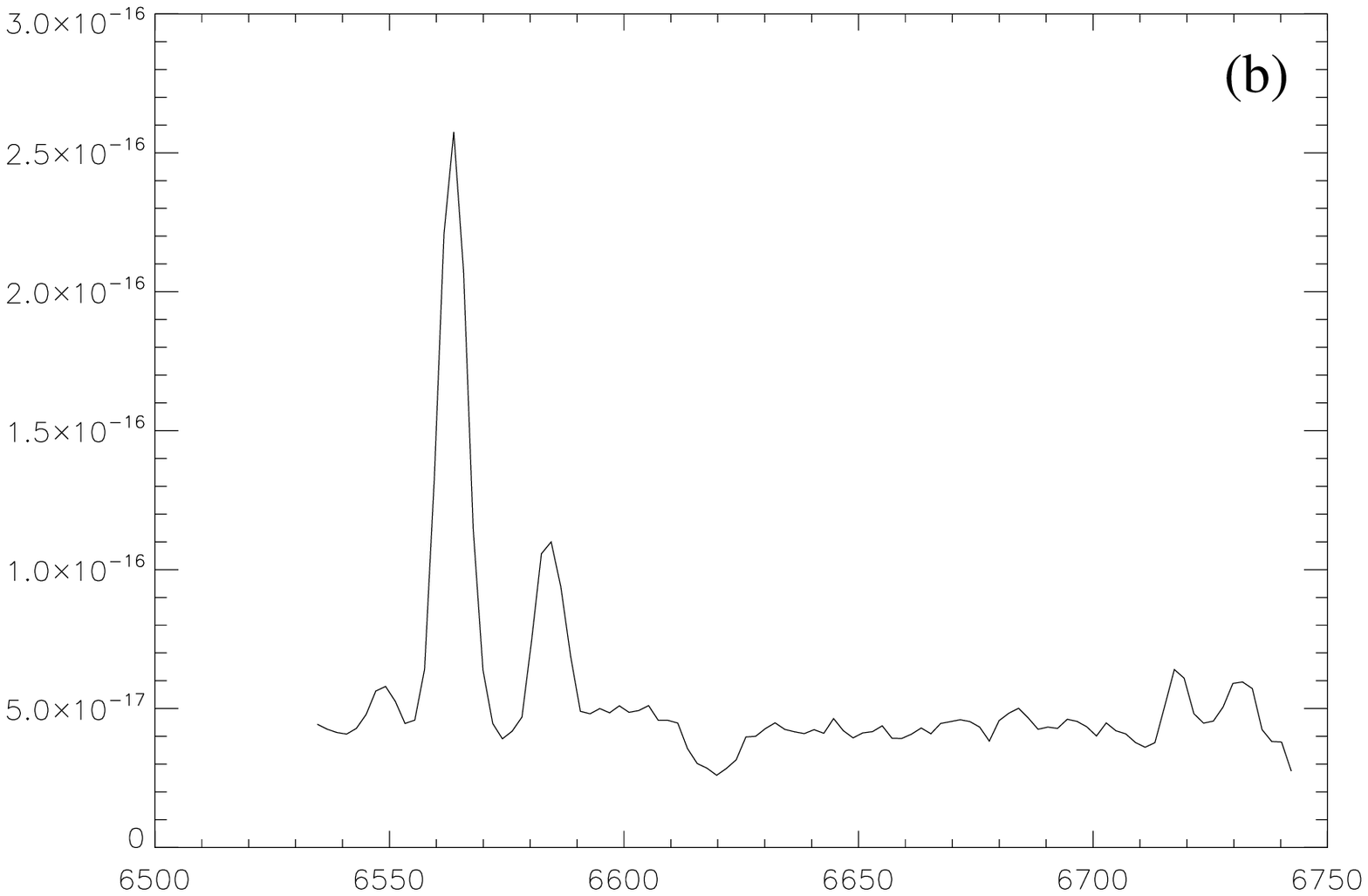}

\end{center}
\caption{Spectra of the merging companion galaxy, illustrating the
  relative strengths of major
  emission lines (arbitrtary flux units). (a - top) - the H$\beta$ and
  [O\textsc{iii}]4959,5007\AA\ emission lines. (b - bottom) -
  [N\textsc{ii}]6549\AA, H$\alpha$, [N\textsc{ii}]6583\AA\ and the
  [S\textsc{ii}]6717,6734\AA\ doublet.
\label{Fig: evil_chick_spectra}}
\end{figure}

Our IFS data include flux from the western edge of this galaxy, and
confirm that it lies at a very similar redshift ($z = 0.2298 \pm 0.0002$) to
PKS1932-46 itself ($z = 0.2307 \pm 0.0002$), i.e. with a rest-frame velocity shift
of $\sim220\pm 70 \rm km\,s^{-1}$ between the two sources.  There are some
indications of a 
velocity gradient across the galaxy, but
the narrow [O\textsc{iii}]5007 line widths typically suggest rather
quiescent gas.  Extracted spectra for this galaxy are displayed in
Figure~\ref{Fig: evil_chick_spectra}, illustrating the major emission
  lines: H$\beta$ and the [O\textsc{iii}]5007 doublet, and also the
  [N\textsc{ii}]6549,6583, H$\alpha$ and  [S\textsc{ii}]6717,6734 lines at longer
  wavelengths.   Using these data, we have plotted
  this galaxy on our BPT diagram (Fig.~\ref{Fig: BPT}); the source
clearly lies in the regime populated by star-forming galaxies. 


\begin{figure}
\vspace{5.2 in}
\begin{center}
\includegraphics{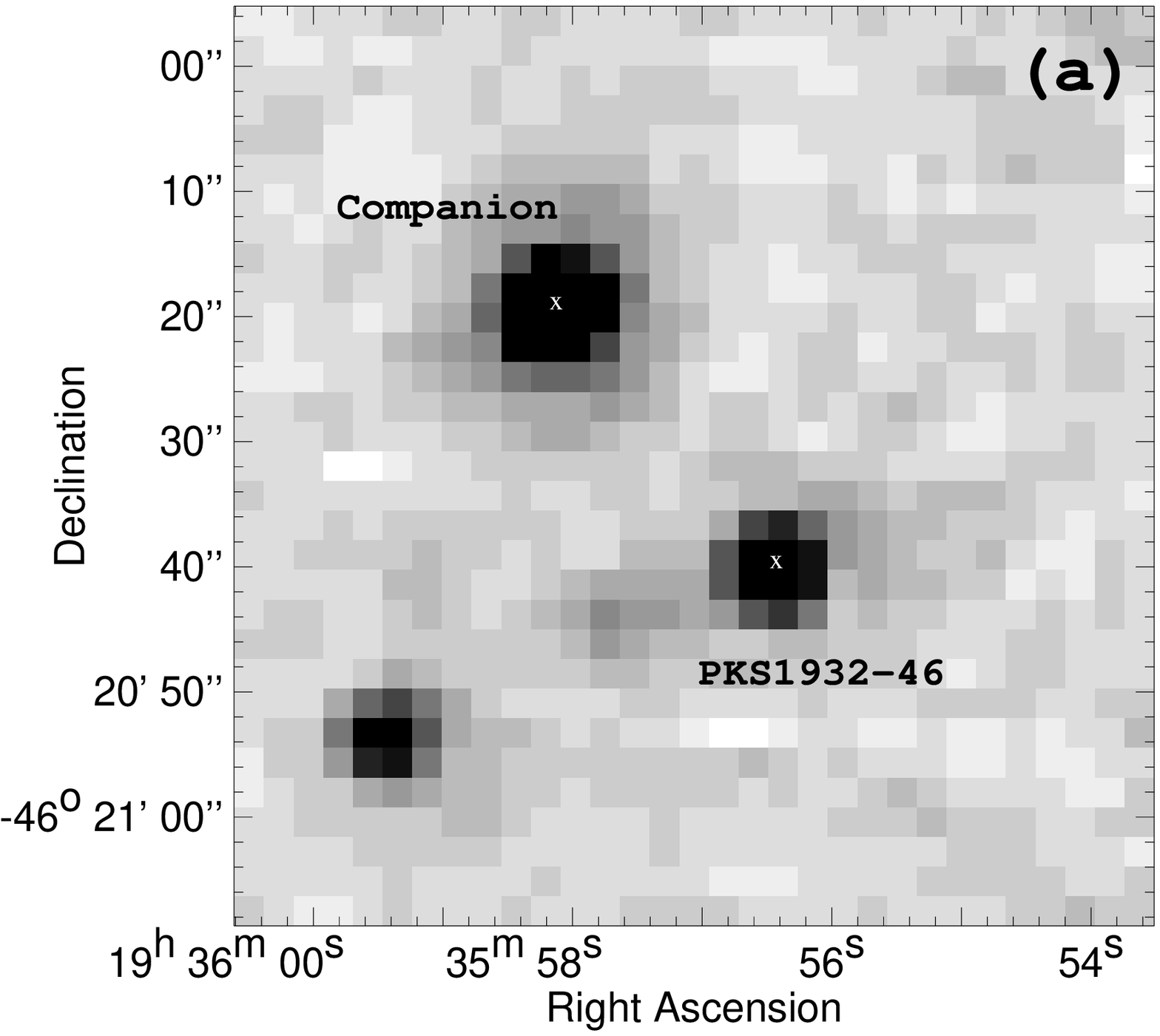}
\includegraphics{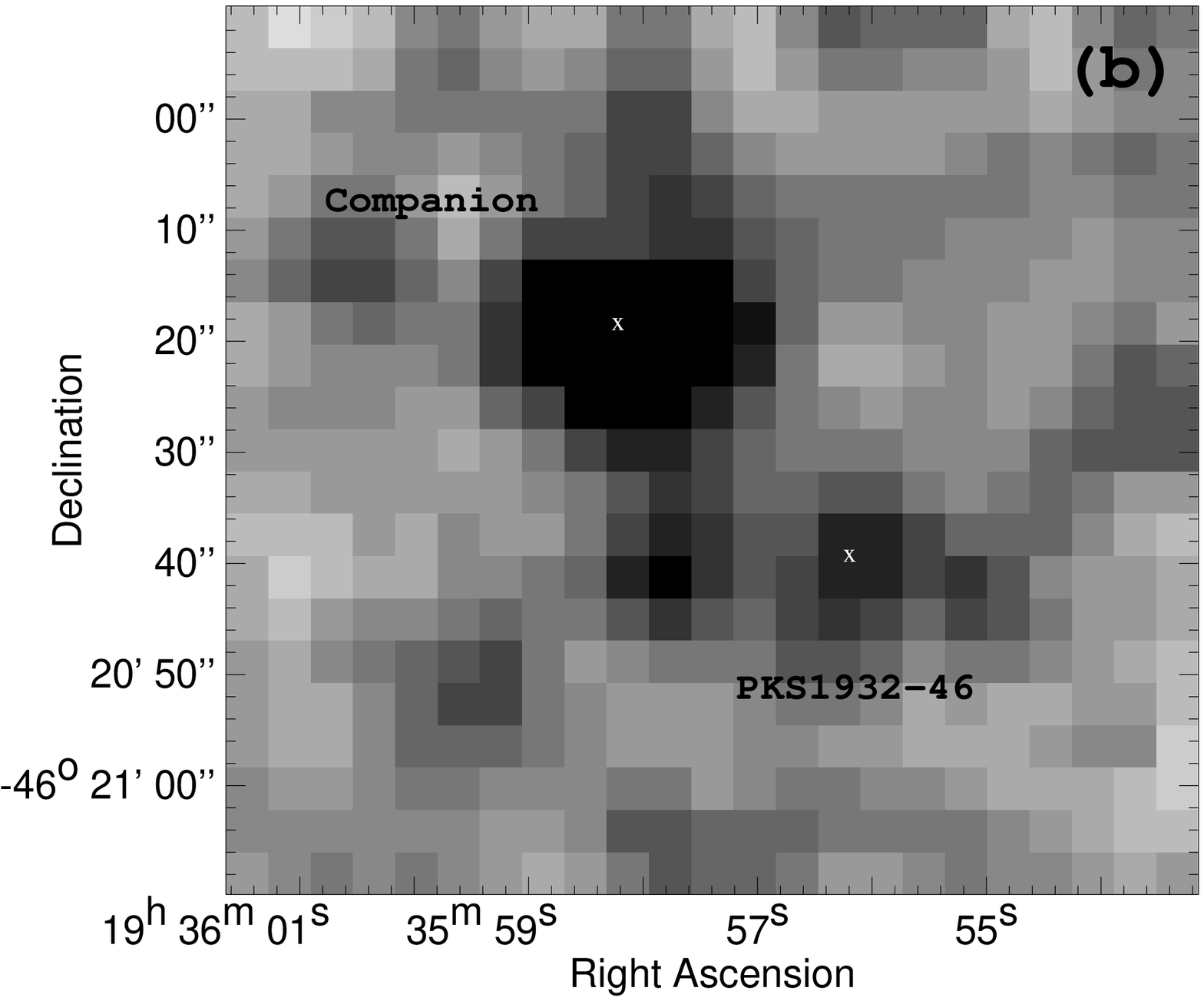}
\end{center}
\caption{24$\mu$m (a - top) and 70$\mu$m (b - bottom) Spitzer MIPS
  images of PKS1932-46 and surrounding objects.  The pixel scales are
  2.54 arcsec per pixel and 4 arscec per pixel for the 24$\mu$m and
  70$\mu$m data respectively.  The PSF FWHM is 6 arcsec for the
  24$\mu$m data, and 18 arcsec for the 70$\mu$m data.
\label{Fig: Spitzer}}
\end{figure}

In addition to very low [N\textsc{ii}]/H$\alpha$ line ratios in the 
companion galaxy spectra, Spitzer MIPS photometry of PKS1932-46 and its 
surrounds reveal that the companion galaxy is  more luminous than the
radio source at 24$\mu$m, and substantially more so at 70$\mu$m,
also suggestive of vigorous star formation in the companion.  These
data are displayed in Fig.~\ref{Fig: Spitzer}, and extracted aperture
fluxes for both the radio galaxy and the companion galaxy are listed
in table 2.  On the basis of the measured MIPS fluxes extrapolated to
the standard IRAS wavebands and the standard formula presented in
Sanders \& Mirabel (1996), we derive
infrared luminosities in our assumed cosmology for the radio source and interacting companion
of $L_{IR} \gta 1.96 \times 10^{44} \rm erg s^{-1}$ ($\gta 5.1 \times 10^{10} \rm
L_{\odot}$) and $L_{IR} \gta 5.0 \times 10^{44} \rm erg s^{-1}$ ($\gta 1.3 \times 10^{11} \rm
L_{\odot}$) respectively, placing the companion galaxy within the
range of luminosities expected for LIRG-type galaxies (Sanders and
Mirabel). In addition, the relatively cool MFIR colours are also
consistent with a starburst scenario.


\section{Discussion}
\subsection{The nature of the host galaxy}

Like most powerful radio sources, the host galaxy of PKS1932-464 is a
massive elliptical.  In conjunction with previous modelling of the
host galaxy stellar continuum, our modelling of the host galaxy
morphology has allowed us to 
determine its mass ($M \sim 1.7-3.2 \times 10^{11}\rm M_{\odot}$)
and size ($r_{\rm eff} \sim 9$kpc), which are fairly modest
for a galaxy of this type. The size of the host galaxy is comparable
to the median size of 3C sources at similar redshifts ($\sim 10$kpc at
$z=0.2$), although
somewhat smaller than the statistical mean ($\sim 12-15$kpc; McLure et al
1999, Roche \& Eales 2000, Inskip et al 2005).  The scatter in the
radio galaxy $K-z$ relation (e.g. Inskip et al 2002c) suggests that a
source at a redshift of $z=0.231$ would 
be expected to have a $K-$band magnitude within the range $\sim 13-15.5$
magnitudes (at these redshifts, the mean and median $K-$band magnitude
is $\sim 14$); our measured $K_S$-band magnitude for PKS1932-46 ($K_S
= 15.3$) places it 
towards the fainter end of this range. Finally, assuming that there is little
intrinsic mass/size variation between galaxies at redshifts of 0.2 to
0.7 (Inskip, Best and Longair 2006), the derived mass for PKS1932-46
lies towards the lower end of the range estimated for 3C galaxies
($\sim 1.5-6.5 \times 10^{11} \rm M_{\odot}$ in our assumed cosmology;
Best et al 1998). We therefore conclude that the central super-massive
black hole is also likely to be proportionately smaller than
the mean value for radio galaxies of similar redshift.

\subsection{The nature of the active nucleus}

\begin{figure}
\vspace{2.4 in}
\begin{center}
\includegraphics{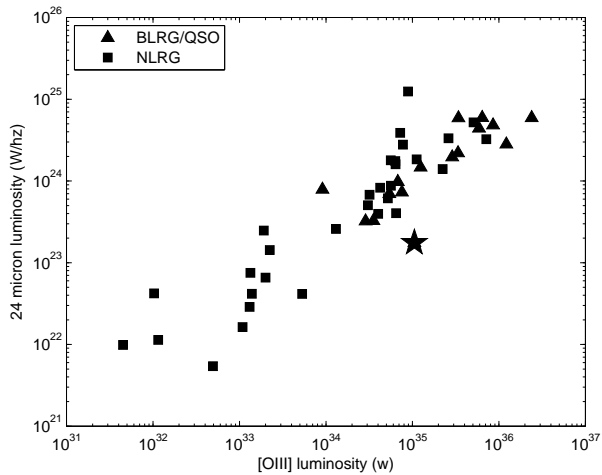}
\end{center}
\caption{Reproduction of Figure 1 
from Tadhunter et al (2007). Square and triangular symbols represent
narrow-line radio galaxies (NLRGs) and quasars/broad-line radio
galaxies (BLRGs) respectively, and PKS1932-46 is marked as a star. 
\label{Fig: puny}}
\end{figure}

In terms of interpreting the nature of the AGN itself, it is
productive to combine the results of different observational studies.

Spitzer MIPS photometry of a complete sample of 2Jy sources (Tadhunter
et al 2007) has been used to address the issue of the dominant dust
heating mechanism in AGN.  The [O\textsc{iii}]5007\AA\ emission line
luminosity (which can be used as a tracer of AGN power, e.g. Rawlings
\& Saunders 1991; Tadhunter et al 1998; Simpson 1998) is strongly
correlated with the mid-IR emission (Fig.~\ref{Fig: puny}),
indicating that AGN illumination is the principal dust-heating
mechanism.  

%
%
%

It is notable that PKS1932-46 lies well below the least-squares fit to
the data on the plot of $24\mu m$ vs. [O\textsc{iii}] luminosities
(Fig.~\ref{Fig: puny}), and also at the lower end of the scatter of
points of  similar [O\textsc{iii}] luminosity.    This can be interpreted in one
of two ways: either (i) the AGN component of the dust
heating is substantially weaker than might be expected for a source of
that [O\textsc{iii}]5007\AA\ emission line luminosity, or (ii) the source
is overluminous in  [O\textsc{iii}]5007\AA\ emission.  However, the
latter explanation seems unlikely given that this object falls well
within the scatter of the main correlation between [O\textsc{iii}]
luminosity and radio power presented by Tadhunter et al (1998) (and is
in fact close to the mean for sources of similar extended radio power).

Observations at other wavelengths can help us to determine whether the
active nucleus is genuinely under-luminous.
Long-slit spectroscopic observations of this source (Holt et al 2007)
have identified a broad emission component in the H$\alpha$ line, with
$L_{H \alpha} \approx 1 \times 10^{41} \rm erg s^{-1}$.  This broad component is 
significantly weaker than that detected in the spectra of other BLRGs
in the 2Jy sample or low redshift 3C BLRGs (typically $5 \times
10^{42} \rm erg s^{-1}$ to several $10^{44} \rm erg s^{-1}$ in our
assumed cosmology; Osterbrock, Koski \& Phillips 1976).  
Using the equations of Greene and Ho (2005)
relating $H{\alpha}$ luminosity and FWHM to black hole mass, the
observed broad $H{\alpha}$ properties suggest a virial mass of
$M_{BH} \sim 1.7 \times 10^7 \rm M_{\odot} \pm 25\%$.  However, based on our
derived bulge mass for PKS1932-46, we would expect the central SMBH to
have a substantially larger mass of $2$ to $7.5\times 10^8 \rm M_{\odot}$ (McLure \& Dunlop
2001, Marconi \& Hunt 2003), appropriate for an $\sim L\star$ galaxy.
This discrepancy suggests that either the 
central quasar itself is underluminous, or that a significant fraction
of the broad emission is blocked from view.

At X-ray wavelengths, ROSAT X-ray observations of PKS1932-46 (Siebert
et al 1996) showed that this source has an X-ray luminosity at
0.1-2.4keV of $6 \times 10^{43} \rm erg\,s^{-1}$ (assuming $H_0 = 50
\rm km\,s^{-1}Mpc^{-1}$ and $q_0 = 0.0$), which corresponds to a
luminosity of $3.3 \times 10^{43} \rm erg\,s^{-1}$ in
our chosen cosmological model.  This value is below the median average ($4.6 \times 10^{43} \rm erg s^{-1}$)
for FRII BLRGs (Siebert et al 1996; see also Fabbiano et al 1984,
Sambruna et al 1999). 
%
%
%
%
Although the X-ray emission is likely to be predominantly due to the
AGN, the presence of a hot x-ray emitting halo associated with the
host galaxy or a group/cluster environment can also contribute to the
observed X-ray fluxes at some level.   
The observed X-ray luminosities (corrected for absorption) therefore represent
upper-limits on the total X-ray emission from the AGN.

\begin{figure}
\vspace{2.25 in}
\begin{center}
\includegraphics{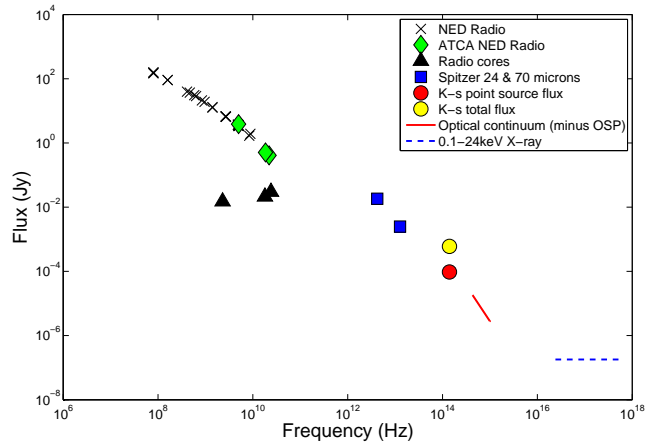}
\end{center}
\caption{Spectral energy distribution for PKS1932-46. Radio data
  points are from NED and ATCA observations. The optical continuum is
  the best-fitting power-law contribution to the galaxy continuum
  emission after deduction of the old stellar population (Holt et al
  2007). The X-ray data is for the average X-ray flux measured by
  Siebert et al (1996). Other data points are as presented in this paper.
\label{Fig: SED}}
\end{figure}

We can also consider these multi-wavelength data as a whole.
Fig.~\ref{Fig: SED} displays the observed-frame SED for PKS1932-46,
from radio through to X-ray frequencies.  In
their modelling of radio source host galaxy stellar populations for PKS1932-46, 
Holt et al (2007) identified the
presence of a power law component of the form $F_{\nu} \propto
\nu^{-\alpha}$ with $\alpha = 2.23$ alongside the continuum
emission from the old stellar population. There is some uncertainty in
the slope of the optical power-law; although the continuum modelling of
Holt et al (2007) is consistent with a combination of old stellar
population and power law components, the addition of younger stellar
populations and various degrees of reddening to the modelling leads to a range of acceptable power-law
fits with spectral
indices of $\alpha \gta 1.4$  (c.f. $\alpha = 
2.23$ for old stellar population models only) and reddening of the
stellar populations of $0 < E(B-V)
< 0.8$.  Assuming that the AGN
is responsible for the bulk of the 24$\mu$m emission and the 16\%
nuclear point source contribution in the K-band, the  modelled IR--optical AGN
contribution is a power-law with $\alpha = 1.4 \pm 0.15$, similar 
to the minimum modelled optical spectral index.  This SED slope is redder
than those measured for  low redshift quasars (Simpson and
Rawlings 2000), 
but certainly bluer than heavily extinguished objects such
as 3C41 (Simpson and Rawlings 2000) and PKS1549-79 (Holt et al 2006).

In addition to the evidence obtained at other wavelengths, the broad H$\alpha$ 
emission detected by VM98 and Holt et al (2007) is also weak.  Taken
together, these results all suggest that PKS1932-46 is powered by an
underluminous AGN when compared with similar 2Jy sources.  
Interestingly, while the AGN appears to be underluminous at many
wavelengths, the source remains powerful in terms of its 
narrow line emission from the nuclear regions. 
One means of explaining the discrepancy between
the observed [O\textsc{iii}] and 24$\mu$m luminosities is in terms of
the source geometry. If the bulk
of the 24$\mu$m emission originates from the torus surrounding the
active nucleus, a smaller than average covering factor for the torus
relative to the NLR would lead to lower than expected 24$\mu$m
emission.  However, this would not explain the underluminous emission
at X-ray and optical wavelengths.  Most plausibly, AGN variability over timescales of $\lta
10^4$ years could account for the lower
than average nuclear luminosities at all wavelengths. Given the differences in light
travel times for the torus and NLR, it is plausible that the central
quasar is currently in a lower state of activity, whereas the luminous
[O\textsc{iii}] and radio emission reflect an epoch in the recent past
when the AGN was more active.

Overall, we conclude that the AGN within PKS1932-46 is most likely to
be powered by an AGN that is currently significantly less luminous
than the AGN in most other sources in the wider 2Jy
sample of similar redshift, and that it is likely that the power-output of the AGN has
recently decreased.



\subsection{EELR properties}

The EELR of PKS1932-46 displays a diverse variety of features.  In
brief:
\begin{enumerate}
\item{} The extended ionised structures display a variety of ionisation
mechanisms, with evidence for ionisation by the AGN, young stars and
possibly also shocks with photoionising precursor regions. AGN
photoionisation has a more obvious impact on the regions along or
close to the radio structures, and in the vicinity of the AGN, while
stellar ionisation becomes more significant at greater distances
from the AGN ionisation cone.  
\item{} Line emission is produced over a wide range of velocities
across the EELR (redshifted by $\sim 200-300\rm km\,s^{-1}$ in the
western radio lobe, and blueshifted by a similar amount for the
star forming material extending northwards from the host galaxy).
However,  the narrow line widths generally reflect fairly quiescent
gas kinematics.  Close to the host galaxy (i.e. within a projected
distance of $\sim 20$kpc from the AGN), the system  displays multiple
velocity  structures along the line of sight.  The broadest line
widths (up to $\sim 1000\rm km\,s^{-1}$) are  also observed in this
region.   
\end{enumerate}

The {\it lack} of disturbed kinematics in the outer regions of the EELR of PKS1932-46 is
one of the noteworthy features of this source. Given that this
source displays very little clear-cut evidence for shocks in the
regions probed by our IFS data, the large extent of the kinematically
quiescent material in the outer regions of the EELR is somewhat
anomalous.   
Overall, the emission properties at large distances from the AGN are more typical of the
photoionised material observed in less extensive low-redshift EELRs
(Inskip et al 2002b), but also reminiscent of the kinematically
quiescent material in the extensive gaseous haloes observed at high redshift (e.g. Reuland et
al 2007).    
Understanding how this complex system came into being may shed light
on the formation of higher redshift EELRs, and the causes of the
observed evolution in their properties. 
In their spectroscopic study of the PA $-9^{\circ}$ material,
VM05 suggested that the best explanation for the properties
of this EELR was  a series of compact, star-forming objects
associated with the tidal debris of a merging system.  Indeed, the narrowness of the observed emission
lines is consistent with those measured in star-forming objects. We concur with
this explanation for the various knots extending across north/south,
which appear to form a large-scale linear structure in the foreground
of the host galaxy.  

However, for the quiescent
material in the western radio lobe (which does not display signs of
star formation) it may be less likely that the emitting material is
self-gravitating. Assuming simple
clouds of approximately 10pc in size, the material would be expected to dissipate over a relatively
short timescale ($\sim 10^4$ years; e.g. Fabian et al 1987), placing
the persistence of the EELR in question. While pressure-equilibrium
with the hot-phase IGM in the galaxy halo could act as a confinement
mechanism in this scenario, alternative descriptions of EELR clouds
allow for longer lifetimes for the emitting regions, e.g. the
overpressured mixed-medium clouds described by Robinson et al (2002). 

Overall, a tidal or outflow origin for the EELR material
(possibly associated with recent star formation) is a very
plausible means of explaining its extreme size.  
\subsection{The local environment - an interacting group?}

%

Redshift evolution is observed in a great many different aspects of the
EELRs associated with powerful radio galaxies, and in many ways
PKS1932-46 bears more resemblance to high-redshift systems than its
low redshift counterparts.  
%
It is plausible  that the similarities between the properties of PKS1932-46 and 
higher redshift systems could be explained by the nature
of its local environment.  
%
%
%
%

Our $K_S$-band
imaging of PKS1932-46 reveals a large number of faint features in the surrounding
field, as well as several bright objects.  The large galaxy to the
northeast of PKS1932-46 is clearly a very disturbed system, displaying
multiple nuclei, tidal tails and evidence for significant ongoing star
formation.  This companion galaxy lies at an identical redshift to the
star-forming material extending north-south across PKS1932-46.  
However, the richest cluster environments are ruled out by the
observed X-ray luminosity of PKS1932-46, which does not allow for a
strong contribution to the X-ray emission from a hot IGM.

We suggest that PKS1932-46 is a member of an interacting galaxy group,
and that a large proportion of the EELR material may originate 
from the tidal debris of previous interactions (plausibly ejected
giant molecular clouds in the case of the star forming regions), either between
PKS1932-46 and the companion galaxy, or from a 
previous merger event. 
Such events, likely to be more common at
higher redshifts, may plausibly also 
have been responsible for triggering the current radio source
activity (e.g. Heckman et al 1986, Wilson \& Colbert 1995), and can
almost certainly explain why this source displays 
such dramatic features in its EELR.   It is interesting to note that
although starbursts and AGN activity may represent two major
manifestations of mergers and interactions (e.g. Springel et al 2005), they are separated within
this group: nuclear activity is dominant in the radio galaxy PKS1932-46, whereas the
dominant star formation activity in this group is associated with the
companion galaxy.

The nature of the surrounding environment also has implications for
our interpretation of the observed galaxy and EELR
properties; specifically, increasing the filling factor of line-emitting clouds can
have several  
knock-on effects for a source such as this. A source with a specific
AGN/jet power would be observed to have a larger radio luminosity when
existing in a higher-density environment (e.g. Kaiser, Dennett-Thorpe
\& Alexander, 1997) than it would otherwise.  Further, boosting of the
observed [O\textsc{iii}] line emission due to other processes would
mean that the measured line luminosity would no longer be indicative
of the intrinsic AGN/jet power of the source.  It is plausible that
both of these factors are at work in the case of PKS1932-46, and that
this may also be the cause of its departure from the correlation
between the [O\textsc{iii}] and mid- to far-IR luminosities in the
2Jy sample.

\section{Conclusions}

Our multiwavelength observations of PKS1932-46 have led to the
following results:

\begin{enumerate}
\item[$\bullet$]{} The host galaxy is an $\sim M_{\star}$ elliptical
  containing a moderately powerful AGN.  Although weak broad line
  emission has been observed from this source, the AGN luminosity is
  likely to be relatively low for a source of this radio power.
  The observed BLRG features suggest that any obscuration of the
  nuclear regions is moderate at best, and that the AGN is
  intrinsically weaker than average for the sample, and thus not all
  powerful FRII radio sources are associated with 
  highly luminous quasar nuclei.  Further, it is probable that the
  power-output of the AGN has dropped significantly over the past
  $10^4$ years.
\item[$\bullet$]{} The extensive emission line region surrounding this
  galaxy consists of kinematically quiescent, AGN-photoionised material
  within the radio lobe, and a band of star-forming knots extending
  north-south, lying perpendicular to the radio axis and offset by
  velocities of a few hundred $\rm km\,s^{-1}$. We observe evidence
  for a gradient in the EELR ionisation mechanism, such as that
  observed in higher redshift sources.
\item[$\bullet$]{} PKS1932-46 lies in a group environment, which is
  dominated by the merging galaxy to the north east.  Unusually it is
  this companion object, rather than the radio source host galaxy,
  which is undergoing the bulk of the star formation activity within
  the group.  However, it is likely that past/ongoing interactions
  within the group may be
  responsible for both the triggering of the radio source, and also
  the richness of the spectacular EELR.
\end{enumerate}

\section*{Acknowledgments}
KJI acknowledges support from a PPARC research fellowship, JH a PPARC
  PDRA, and DD a PPARC research studentship. The work of MV-M  has
  been supported by the Spanish Ministerio de Educaci\'on y Ciencia
  and the Junta de Andaluc\'ia through the grants AYA2004-02703 and TIC-114.
This work is based in  part on observations made with the Spitzer Space Telescope, which
  is operated by the Jet Propulsion Laboratory, California Institute
  of Technology under a contract with NASA.  The IFS data published in this paper have been reduced using VIPGI,
designed by the VIRMOS Consortium and developed by INAF Milano; KJI  
would particularly like to thank Paolo Franzetti for guidance in the
use and reliability of the VIPGI software.   We also thank the
ESO technical and support staff for indulging our request for the use of SOFI in the
small field mode.  This research has made use of the NASA/IPAC
  Extragalactic Database (NED) which is operated by the Jet Propulsion
  Laboratory, California Institute of Technology, under contract with
  the National Aeronautics and Space Administration.  We thank the
  referee, Alan Stockton, for his careful consideration of the manuscript.

\label{lastpage}

\end{document}